\documentclass[11pt,a4paper]{article}
\usepackage{jcappub}

\usepackage{bm}
\usepackage{graphicx} 
\usepackage{hyperref}
\usepackage{mathrsfs}

\usepackage{latexsym}
\usepackage{amssymb}
\usepackage{epsfig}
\usepackage{amsmath}
\usepackage{float}
\usepackage[utf8]{inputenc}

\usepackage{color}
\usepackage{xcolor,colortbl}

\newcommand{\AEST}{\text{AeST}}
\newcommand{\Poisson}[2]{\ensuremath{ \left\{ #1, #2\right\} }}   
\newcommand{\grad}{\ensuremath{\vec{\nabla}}}   
\newcommand{\chis}{\ensuremath{\chi_s}}

\newcommand{\wt}{\ensuremath{\tilde{w}}}
\newcommand{\GN}{\ensuremath{G_\text{N}}}
\newcommand{\rC}{\ensuremath{r_C}}
\newcommand{\rM}{\ensuremath{r_M}}
\newcommand{\muf}{{f}}
\newcommand{\PsiM}{\Psi}
\newcommand{\cs}{ c_s}
\newcommand{\cvis}{ c_{\rm vis}}
\newcommand{\Ab}{\overline{A}}

\newcommand{\Qsmall}{{q}}

\newcommand{\Qcal}{{\cal Q}}
\newcommand{\Qcalb}{\overline{{\cal Q}}}
\newcommand{\nablab}{\bar{\nabla}} 
\newcommand{\kpc}{{\rm kpc}}
\newcommand{\km}{{\rm km}}
\newcommand{\Mpc}{{\rm Mpc}}
\newcommand{\Scal}{{\cal S}}
\newcommand{\Scalb}{\overline{{\cal S}}}
\newcommand{\Tcal}{{\cal T}}
\newcommand{\Ocal}{{\cal O}}
\newcommand{\Jcal}{\ensuremath{{\cal J}}}   
\newcommand{\Rcal}{{\cal R}}
\newcommand{\Gcal}{{\cal G}}

\newcommand{\Wcal}{{\cal W}}
\newcommand{\Zcal}{{\cal Z}}
\newcommand{\Ham}{ { \cal H}}

\newcommand{\Ycal}{{\cal Y}}

\newcommand{\Mcal}{{\cal M}}

\newcommand{\lambdad}{\lambda_{D}}
\newcommand{\taub}{\overline{\tau}} 
 
\newcommand{\Kcal}{{\cal K}} 
\newcommand{\Lcal}{{\cal L}} 
\newcommand{\Kcalb}{\overline{{\cal K}} }
\newcommand{\rhob}{\overline{\rho} }
\newcommand{\Pb}{\overline{P} }
\newcommand{\rhoM}{{\rho_m}}

\newcommand{\cad}{c_{\rm ad}}  
\newcommand{\Pinad}{\Pi_{\rm nad}}

\newcommand{\Nb}{\overline{N}}

\definecolor{orange}{rgb}{1,0.5,0}
\definecolor{darkorange}{rgb}{0.69,0.33,0.13}
\definecolor{fidcol}{rgb}{0.7,0,0}

\newcommand{\dd}{\mathrm{d}}
\newcommand{\di}{\mathrm{i}} 
\newcommand{\de}{\mathrm{e}}

\allowdisplaybreaks

\allowdisplaybreaks

\makeatletter
\gdef\@fpheader{Published in JCAP11(2024)040}
\makeatother

\begin{document}

\title{Relativistic Khronon Theory \\in agreement with \\Modified Newtonian Dynamics and Large-Scale Cosmology}

\author[a]{Luc \textsc{Blanchet}}\emailAdd{luc.blanchet@iap.fr}
\author[b,c]{Constantinos \textsc{Skordis}}\emailAdd{skordis@fzu.cz}

\affiliation[a]{$\mathcal{G}\mathbb{R}\varepsilon{\mathbb{C}}\mathcal{O}$, 
	Institut d'Astrophysique de Paris, UMR 7095, CNRS, Sorbonne Universit{\'e},
        98\textsuperscript{bis} boulevard Arago, 75014 Paris, France}
\affiliation[b]{CEICO, Institute of Physics of the Czech Academy of Sciences, Na Slovance 1999/2, 182 21, Prague, Czech Republic}
\affiliation[c]{Department of Physics, University of Oxford, Denys Wilkinson Building, Keble Road,   Oxford OX1 3RH, UK}

\abstract{
We propose an extension of General Relativity (GR) based on a space-time foliation by three-dimensional space-like hypersurfaces labeled by the Khronon scalar field $\tau$. We show that this theory (i) leads to modified Newtonian dynamics (MOND) at galactic scales for stationary systems; (ii) recovers GR plus a cosmological constant in the strong field regime; (iii) is in agreement with the standard cosmological model and the observed cosmic microwave background anisotropies at linear cosmological scales, where the theory reduces to a subset of the generalized dark matter (GDM) model. We compute the  second order action on a Minkowski background and show that it contains the usual tensor modes of GR and a scalar degree of freedom with dispersion relation $\omega=0$. We find that the  deconstrained Hamiltonian is bounded from below for wavenumbers larger than $\sim 10^{-31}\,\text{eV}$ and unbounded for smaller wavenumbers. 
}

\arxivnumber{2404.06584v2}

\maketitle

\section{Introduction}
\label{Sec:Introduction}

More than forty years ago Milgrom~\cite{Milg1, Milg2, Milg3} proposed an empirical formula, dubbed MOND for ``MOdified Newtonian Dynamics'', able to fit with astonishing precision the rotation curves of galaxies and the Tully-Fisher relation between the baryonic mass and the asymptotic rotation velocity of galaxies. The MOND formula permits to resolve a number of challenges faced by the standard cosmological model $\Lambda$-CDM (Cold Dark Matter plus a cosmological constant $\Lambda$) at galactic scales~\cite{FamMcG12}. It suggests a modification of fundamental physics, notably the gravitational sector described by general relativity (GR), in a regime of low accelerations, below a certain critical acceleration scale $a_0$, empirically measured at the value $a_0\simeq 1.2\,10^{-10}\,\mathrm{m}/\mathrm{s}^2$.  

The MOND formula has prompted the construction of several extensions of GR~\cite{BekensteinMilgrom1984,Bekenstein1988}, which purport to fit cosmological observations without the need of dark matter. A well-studied example is the Tensor-Vector-Scalar (TeVeS) theory of Bekenstein and Sanders~\cite{Sand97, Bek04, Sand05}, which extends GR with a time-like vector field (akin to Einstein-Aether theory~\cite{JM00, JM04}) and a scalar field; see~\cite{Skordis2009} for a review. Einstein-Aether theories, originally motivated by the phenomenology of local Lorentz invariance violation~\cite{JM00, JM04}, were extended to lead to the MOND formula at the regime of galaxies~\cite{ZFS07, HZL08}. Further GR extensions with a MOND limit were proposed in~\cite{Milgrom2009, BDgef11, DEW11, BM11, Sanders2011, Mendoza:2012hu, Woodard2014, K15, BH15a, Hossenfelder2017, Burrage:2018zuj, Milgrom:2019rtd, DAmbrosio:2020nev, Kading:2023hdb} and the cosmological and astrophysical predictions of several of these theories have been extensively studied~\cite{SMFB06, BourliotEtAl2006, DodelsonLiguori2006, LBMZ08, Sk08, Zu10, XuEtAl2014, DaiStojkovic2017, ZlosnikSkordis2017, TanWoodard2018}. The recent Aether-Scalar-Tensor ($\AEST$) theory proposed by Skordis and Zlosnik~\cite{SZ21} based on~\cite{SkordisZlosnik2019}, has been the first example of a GR extension reproducing the MOND formula in galaxies and simultaneously being in agreement with the standard cosmological model $\Lambda$-CDM, \emph{i.e.} predicting the full spectrum of anisotropies of the cosmic microwave background (CMB).\footnote{A different class of MOND theories attempt to modify the properties of dark matter itself, rather than the gravitational law. An example is the dipolar dark matter model~\cite{BL08, BL09} which is also in agreement with the standard cosmological model $\Lambda$-CDM in first order cosmological perturbations.} 

A simple extension of GR leading to MOND behaviour in galaxies was proposed by Blanchet and Marsat~\cite{BM11}, hereafter BM theory, as well as an independent variant by Sanders~\cite{Sanders2011}, based on only two dynamical fields, the metric $g_{\mu\nu}$ and a scalar field $\tau$ called the Khronon, and labeling a family of space-like three-dimensional hypersurfaces. All ordinary matter fields in this theory are universally coupled to the metric. However, in the local freely falling frame associated with the matter fields, the presence of the Khronon field induces an effective violation of the local Lorentz invariance (LLI). The BM theory was inspired by the Ho{\v{r}}ava-Lifshitz approach~\cite{Horava2009, Blas2009, Blas2011} for a possible violation of LLI and a completion of GR in a (power-counting) renormalizable theory at high energy. However, its motivation is quite different from that of~\cite{Horava2009, Blas2009, Blas2011}, as it is not concerned with the problem of quantum gravity, and postulates the violation of LLI at low energy, in the MOND regime for weak gravitational fields. This idea has been generalized by Bonetti and Barausse~\cite{Bonetti2015} and the BM theory has recently been further discussed by Flanagan~\cite{Flanagan23}. 

The main drawback of the BM theory, that concerns us here, is that it does not explain the effects commonly attributed to  dark matter, which occur at large scales in cosmological perturbations. We therefore extend the BM theory by adding, in a quite natural way, a kinetic term for the Khronon scalar field. Following~\cite{Scherrer2004, Arkani2004, Arkani2007} we know that a large class of kinetic $k$-essence terms are able to mimic dark matter in perturbations around an homogeneous and isotropic Friedman-Lema\^{i}tre-Robertson-Walker (FLRW) background. Hence, we shall prove that with this crucial addition the model, for an appropriate choice of the Khronon kinetic term, is in agreement with the standard cosmological model at large scales (to first order in cosmological perturbation), and in particular retrieves the full observed spectrum of anisotropies of the CMB, while still providing the relevant MOND limit. 

More precisely, we show that the Khronon equations can be recast into the framework of the generalized dark matter (GDM) model~\cite{Hu:1998kj, Kopp:2016mhm}. 
The Khronon behaves like a GDM fluid with time-dependent equation of state in the background, and with zero viscosity and non-zero, $k$-dependent sound speed in linear perturbations. Using the GDM framework we investigate the allowed constraints on cosmology and the CMB (namely the cosmological dark matter which should be close enough to a ``pure dust'' model) which are consistent with recovering the MOND phenomenology at galactic scales. In particular we find that a kinetic term for the Khronon based on the Dirac-Born-Infeld (DBI) form of the action reproduces in a natural way all observations. 

Finally, we also show that the theory recovers GR plus a cosmological constant in the strong field, high acceleration regime, and in particular has the same parametrized post-Newtonian (PPN) parameters as GR for tests in the Solar System. 

The paper is organized as follows. In section~\ref{Sec:Action} we describe the covariant action of the theory and derive the field equations. We discuss in appendix~\ref{Sec:adapted} an equivalent formulation of the theory in the coordinate system adapted to the Khronon, that is, when the time coordinate is identified with the Khronon field itself. In section~\ref{Sec:NR_limit} we construct the nonrelativistic slow motion limit of the theory and discuss how Newtonian behaviour at high-acceleration (and GR in the strong-field high-acceleration limit) is reached and how MOND behaviour at the low-acceleration limit is obtained. In section~\ref{Sec:cosmo} we show how some specific choices for the new term in the action that we propose in the present work leads to cosmology in agreement with observations of large scale structure and the CMB radiation. In section~\ref{Sec:stability} we expand the action to second order in perturbations on a Minkowski background, determine the normal modes and compute the deconstrained Hamiltonian (in particular we find that the theory has the same tensorial modes as GR). Section~\ref{Sec:discussions} ends with a short discussion of the Ho{\v{r}}ava khronometric theory and presents our conclusions.

We use a $(-+++)$ metric signature convention and curvature convention of MTW~\cite{MTW}. We denote general coordinates by $x^\mu$ which are taken to have dimensions of length, such that, for a time coordinate $t$ we have $x_0 = c t$ with $c$ being the speed of light. Partial $\partial_\mu$ and covariant derivatives $\nabla_\mu$ are with respect to $x^\mu$ and have dimensions of inverse length. We denote by $\gamma_{ij}$ the Euclidean metric in general coordinates (reserving $\delta_{ij}$ for the Kronecker symbol in cartesian coordinates), and by $\grad_i$ the covariant derivative associated with $\gamma_{ij}$. We use the vector calculus notation $\grad_i \leftrightarrow \grad$ and $\grad u \cdot \grad v \equiv \grad_i u  \grad^i v = \gamma^{ij} \grad_i u  \grad_j v$ for any scalars $u$ and $v$. In cosmology we denote by $\gamma^{\kappa}_{ij}$ a spatially homogeneous and isotropic metric of spatial curvature $\kappa$, hence $\kappa =0$ corresponds to a spatially flat Universe, so that $\gamma^{0}_{ij}=\gamma_{ij}$, and  $\kappa>0$ ($\kappa<0$) to a positively (negatively) spatially curved Universe. We have $\gamma^{\kappa}_{ij} = \gamma_{ij}(1+\frac{\kappa}{4}\,|\vec{x}|^2)^{-2}$ in a spherical coordinate system.

\section{Action and field equations}
\label{Sec:Action}

\subsection{General definitions}

The dynamical degrees of freedom of the theory comprise the metric $g_{\mu\nu}$, the Khronon scalar field $\tau$ which is assumed to have units of time, and the ordinary matter fields $\PsiM$. The Khronon defines a spatial foliation through the orthonormal unit-timelike vector field $n^\mu$ defined via
\begin{align}
\label{nmu_def}
	n_\mu &= - \frac{c}{\Qcal} \nabla_\mu \tau\,,
\end{align}
with the useful scalar quantity 
\begin{align}
\label{Qcal_def}
\Qcal &\equiv c\sqrt{- g^{\mu\nu} \nabla_\mu \tau \nabla_\nu \tau}\,.
\end{align}
Both $n_\mu$ and $\Qcal$ are dimensionless. By construction one can see that $n^\mu n_\mu  = -1$ and that $n^\mu$ is future pointing, $n^0 > 0$. The vector $n^\mu$ serves to define the (covariant) spacelike acceleration of the congruence as\footnote{In~\cite{BM11, Sanders2011} the acceleration is denoted $a^\mu$; however, here we reserve $a$ for the scale factor in cosmology.} 
\begin{align}\label{Amu_def}
	A_\mu =  c^2 n^\nu \nabla_\nu n_\mu = - c^2 q_{\mu}^{\;\;\nu}  \,\nabla_\nu \ln \Qcal \,,
\end{align}
where the factor of $c^2$ ensures that it has units of acceleration, and where the projector onto the family of spatial hypersurfaces is denoted as 
\begin{align}
q_{\mu}^{\;\;\nu} \equiv \delta_\mu^{\nu}  + n_\mu n^\nu\,,
\end{align}
so that $n^\mu A_\mu = 0$. 
Note that the spatial foliation~\eqref{nmu_def}--\eqref{Amu_def} is invariant by arbitrary reparametrization of the Khronon field: $\tau \rightarrow \tau'=F(\tau)$. However the kinetic term for the Khronon, eq.~\eqref{K_def} below, will break the reparametrization invariance of the model.  

Both in the BM theory~\cite{BM11} and in the Sanders variant~\cite{Sanders2011}, the acceleration~\eqref{Amu_def} is used to lead to  MOND behaviour in the nonrelativistic and low acceleration regime. This is achieved through a function $\Jcal(\Ycal)$ of the covariant acceleration~\eqref{Amu_def} squared denoted
\begin{align}
	\Ycal \equiv  \frac{A_\mu A^\mu}{c^4}\,,
\label{Y_def}
\end{align}
that appears in the action functional of the theory. The function $\Jcal$ has the dimension of an inverse squared length and is chosen such that when $\Ycal$ is large (the high acceleration limit) GR is recovered, while when $\Ycal$ is small (the low acceleration limit) and in addition we assume slow-motion quasi stationary systems, MOND behaviour is reproduced. We discuss the details of how this happens, as well as choices of $\Jcal(\Ycal)$ which have this property, in section~\ref{Sec:NR_limit}.

\subsection{Covariant action and field equations}
\label{Sec:covariant}

The covariant action of the theory is
\begin{align}
	S =  \frac{c^3}{16\pi G}\int \dd^4x \,\sqrt{-g} \,\Bigl[ R - 2 \Jcal(\Ycal) + 2 \Kcal(\Qcal) \Bigr] 
 + S_m\left[\PsiM,g\right]\,,
\label{K_action}
\end{align}
where $R$ is the four-dimensional Ricci scalar, $g$ is the metric determinant and $G$ the gravitational coupling strength. The matter fields $\PsiM$ are universally coupled to the metric $g_{\mu\nu}$ 
and as such the Einstein equivalence principle holds (we regard the Khronon field as part of the gravitational sector). Moreover,
 in the high acceleration limit GR is recovered and as such, the strong  equivalence principle (SEP) holds there, however, it does not hold at the low acceleration limit which includes MOND and the 
$\mu$-domination regime.

The function $\Jcal(\Ycal)$ appearing in~\eqref{K_action}, where $\Ycal$ is defined by~\eqref{Y_def}, has already been postulated in~\cite{BM11} and~\cite{Sanders2011} and is part of the original BM theory. Our new addition is the function $\Kcal(\Qcal)$ which depends on the scalar field $\Qcal$ defined in~\eqref{nmu_def} and serves as a kinetic term for the Khronon field $\tau$. Following~\cite{SZ21}, our main assumption about this function is that it admits a Taylor expansion around the value $\Qcal=1$. For definiteness we specifically adopt in this section
\begin{align}
	\Kcal(\Qcal) = \mu^2\left(\Qcal -1\right)^2\,,
	\label{K_def}
\end{align}
and leave more general functions having such a Taylor expansion investigated in section~\ref{Sec:cosmo-MOND}. Here the constant $\mu$ has dimensions of inverse length, and such a term in the action plays a vital role in the success of $\AEST$ theory in fitting cosmological observations~\cite{SZ21}. A similar term appears in the case of ghost condensation, which is an extension of GR associated with the breaking of time diffeomorphisms~\cite{Arkani2004} and in the case of shift-symmetric $k$-essence~\cite{Scherrer2004}. In both of these last two cases the relevant term appearing in the function $\Kcal$ is, rather, $\Kcal(X) = \mu^2(X+1)^2$ with $X = c^2 \nabla_\mu \phi \nabla^\mu\phi$ for a generic scalar $\phi$ (using our metric signature conventions), $\phi=\tau$ and $X=-\Qcal^2$ in our case.

The term $\Kcal(\Qcal) = \Kcal(\sqrt{-X})$ in the action \eqref{K_action} belongs to the general class of Horndeski theories~\cite{Horndeski}, as a particular case of the $k$-essence term $L_2^\text{H}\equiv G_2(\phi,X)$. The acceleration squared~\eqref{Y_def} is part of the more general class of DHOST theories, as we have
\begin{align}
	\Ycal = \frac{L_4^{(2)}}{X^2} - \frac{L_5^{(3)}}{X^3} \,,
	\label{DHOST}
\end{align}
where $L_4^{(2)} \equiv \phi^{\nu}\phi_{\mu\nu}\phi^{\rho}\phi^{\mu}_{\rho}$ and $L_5^{(3)} \equiv (\phi^{\mu}\phi^{\nu}\phi_{\mu\nu})^2$ with standard notations~\cite{Langlois19, DeFelice22}. However the action~\eqref{K_action} contains a function of $\Ycal$, involving higher powers of $\Ycal$, see \eqref{lambdaMond} below, and the theory lies outside the class of DHOST theories. 

By varying the action with respect to the metric we obtain the following generalized Einstein field equations
\begin{align}
	G^{\mu\nu} +  \left(   \Jcal -  \Kcal \right)   g^{\mu\nu} 
	- \frac{2}{c^4} \Jcal_\Ycal   A^\mu A^\nu 
	+\left[ \frac{2}{c^2}  \nabla_\rho \left( \Jcal_\Ycal  A^\rho \right) - \Qcal \,\Kcal_\Qcal \right]  n^\mu n^\nu  
	= \frac{8 \pi G}{c^4} \,T^{\mu\nu}\,,
\label{EFE}
\end{align}
where $G^{\mu\nu}=R^{\mu\nu}-\frac{1}{2}R \,g^{\mu\nu}$ is the Einstein tensor, $T^{\mu\nu}=\frac{2}{\sqrt{-g}}\,\delta S_m/\delta g_{\mu\nu}$ is the matter stress-energy tensor, and where $\Jcal_\Ycal\equiv\dd\Jcal/\dd\Ycal$ and $\Kcal_\Qcal\equiv\dd\Kcal/\dd\Qcal$ denote the derivatives with respect to the arguments $\Ycal$ and $\Qcal$. Taking the space-time trace of eq.~\eqref{EFE} we obtain
\begin{align}
	- R + 4 \Jcal - 2 \Ycal \Jcal_\Ycal - 4 \Kcal + \Qcal \Kcal_\Qcal - \frac{2}{c^2} \nabla_\mu \left( \Jcal_\Ycal  A^\mu \right)  
	= \frac{8 \pi G}{c^4} \,T\,.
\end{align}
We define the stress-energy tensor $\Tcal^{\mu\nu}$ of the Khronon field to be such that (by definition)
\begin{align}\label{EFE2}
	G^{\mu\nu}
	= \frac{8 \pi G}{c^4} \Bigl[T^{\mu\nu} + \Tcal^{\mu\nu}\Bigr]\,,
\end{align}
and we read directly from eq.~\eqref{EFE} 
\begin{align}
	\Tcal^{\mu\nu} = \frac{c^4}{8 \pi G} \left\{ - \left(   \Jcal -  \Kcal \right)   g^{\mu\nu} 
	+ \frac{2}{c^4} \Jcal_\Ycal   A^\mu A^\nu 
	+\left[ - \frac{2}{c^2}   \nabla_\rho \left( \Jcal_\Ycal  A^\rho \right)    
	+  \Qcal \,\Kcal_\Qcal \right]  n^\mu n^\nu  \right\}\,.
\label{Tmunutau}
\end{align}
From the contracted Bianchi identity $\nabla_\nu G^{\mu\nu}\equiv 0$ and the equations of motion satisfied by the matter fields $\PsiM$, which imply the covariant conservation of the matter tensor $T^{\mu\nu}$, we infer that the stress-energy tensor of the Khronon must also be covariantly conserved,
\begin{align}\label{nablaTmunutau}
	\nabla_\nu \Tcal^{\mu\nu}=0\,.
\end{align}
Finally we vary with respect to the Khronon field and obtain the covariant conservation law $\nabla_\mu \Scal^\mu = 0$ where the current vector $\Scal^\mu$ reads as\footnote{The Khronon equation $\nabla_\mu \Scal^\mu = 0$ reduces to eq.~(2.8) in the BM model~\cite{BM11} when $\Kcal=0$.}
\begin{equation}
	\Scal^\mu = 
	\frac{2}{\Qcal} \Jcal_\Ycal \,A^{\mu} \,n^{\nu} \nabla_\nu \ln \Qcal + \left[ 
	\frac{2}{\Qcal}  \nabla_\nu \left( \Jcal_\Ycal  \,A^\nu \right)    
	-  c^2 \,\Kcal_\Qcal  \right] n^\mu \,.
\label{Scal}
\end{equation}
As a check of the consistency of the formalism, we can derive the conservation equation of the Khronon stress-energy tensor~\eqref{nablaTmunutau} without reference to the Bianchi identity, \emph{i.e.} directly from the explicit expression~\eqref{Tmunutau} and the Khronon equation $\nabla_\mu \Scal^\mu = 0$.

The formulation of the theory in the adapted coordinate system for which $t=\tau$ (unitary gauge), is relagated to appendix~\ref{Sec:adapted}.

\section{Nonrelativistic limit}
\label{Sec:NR_limit}

\subsection{Post Newtonian expansions for the fields}

We investigate the nonrelativistic (or slow motion) limit of the model, in the case of an isolated matter system (\emph{e.g.} the Milky Way galaxy), and determine under which conditions we can recover the MOND phenomenology in the weak acceleration regime. In this approximation, we introduce two \textit{a priori} different scalar potentials $\phi$ and $\psi$, as well as the ``gravitomagnetic'' vector potential $\zeta_i$, and make the usual post-Newtonian (PN) ansatz on the metric components generated by the isolated system, using the ADM form~\eqref{ADM},
\begin{subequations}\label{PNmetric}
	\begin{align}
		N &= 1 + \frac{\phi}{c^2} + \Ocal(  c^{-4} )\,, \\
		N_{i} &= \frac{4}{c^3}\,\zeta_i +\Ocal( c^{-5} )\,, \label{PNNi}\\
		q_{ij} &= \gamma_{ij}\left(1 - \frac{2\psi}{c^2}\right) + \Ocal( c^{-4} ) \,,
	\end{align}
\end{subequations}
where $\gamma_{ij}$ is a Euclidean metric in general coordinates, and with usual notation for the PN remainder terms $\Ocal( c^{-n})$. The PN ansatz~\eqref{PNmetric} is standard in GR and follows from the leading PN order of the stress-energy tensor $T^{\mu\nu}$ for ordinary matter fields, namely
\begin{subequations}\label{PNTmunu}
	\begin{align}
		T^{00} &= \rhoM c^2 + \Ocal( c^0 )\,, \\
		T^{0i} &= \rhoM v_m^i c + \Ocal( c^{-1} )\,, \\
		T^{ij} &= \rhoM v_m^i v_m^j + p_m \,\gamma^{ij} + \Ocal( c^{-2} ) \,,
	\end{align}
where $\rhoM$, $p_m$ and $v_m^i$ are the mass density, pressure and coordinate velocity of the matter field in the Newtonian approximation. Here $\rhoM$ is the coordinate density, to which $\rho$ [see eqs.~\eqref{matter3d}] reduces in the Newtonian limit, and obeys the ordinary continuity equation
\begin{equation}\label{continuity}	
	\dot{\rho}_m + \grad_i(\rhoM v_m^i)=0\,.
\end{equation}
\end{subequations}
The dot denotes the time derivative and $\grad_i$ is the covariant derivative associated with $\gamma_{ij}$.

In order to justify the PN ansatz~\eqref{PNmetric} for the present theory we must also take into account the Khronon field equation. In particular, we must ensure that the stress-energy tensor $\Tcal^{\mu\nu}$ of the Khronon field given by eq.~\eqref{Tmunutau}, admits the same leading PN behaviour as for the ordinary matter in~\eqref{PNTmunu}. As shown by Flanagan~\cite{Flanagan23} this will be satisfied if we assume that the expansion of the Khronon field is of the type
\begin{align}\label{tausigma}
	\tau = t + \frac{\sigma(\mathbf{x},t)}{c^2} + \Ocal( c^{-4} )\,,
\end{align}
where $\sigma$ is the leading order perturbation. We shall check this point in eq.~\eqref{TcalPN} below. 

Inserting~\eqref{tausigma} into~\eqref{nmu_def} we obtain the PN expansion of the scalar field $\Qcal$ and the foliation's unit vector $n_\mu$ as
\begin{subequations}\label{Qnmunr}
	\begin{align}
	\Qcal &= 1 - \frac{\Xi}{c^2} + \Ocal( c^{-4} )\,,\label{Qnr}\\
	n_0 &= -1 - \frac{1}{c^2}\bigl( \Xi +\dot{\sigma} \bigr) + \Ocal( c^{-4} )\,,\\
	n_i &= - \frac{1}{c}\left(1 + \frac{\Xi}{c^2} \right) \grad_i\sigma + \Ocal( c^{-5} )\,,
\end{align}
\end{subequations}
where we have defined (following the notation in~\cite{Flanagan23})
\begin{equation}\label{Xi}
	\Xi\equiv \phi - \dot{\sigma} + \frac{1}{2}|\grad\sigma|^2\,.
\end{equation}
We obtain in turn the acceleration vector $A_\mu$ defined by~\eqref{Amu_def} as
\begin{subequations}\label{Amunr}
	\begin{align}
	A_0 &= \frac{1}{c} \,\grad \sigma \cdot \grad\Xi + \Ocal( c^{-3} )\,,\\ A_i &= \grad_i\Xi + \Ocal( c^{-2} )\,.
	\end{align}
\end{subequations}

To fully determine the slow motion limit, we need to know the PN order of the functions $\Jcal$ and $\Kcal$ in the action. We find that 
\begin{align}
\Ycal = \frac{1}{c^4} |\grad \Xi|^2 + \Ocal(c^{-5}),
\end{align}
and as we show in~\eqref{lambdaMond} below, in the deep MOND limit $\Jcal  \sim \Lambda - \Ycal + c^2 a_0^{-1} \Ycal^{3/2}  + \Ocal(\Ycal^{2})$, hence, the PN order of this function is $\Jcal= \Ocal(c^{-4})$. This is consistent with the fact that the cosmological constant scales with the MOND acceleration constant $a_0$ like $\Lambda\sim a_0^2/c^4$ and is thus formally $\Ocal(c^{-4})$.

The PN order of the function $\Kcal$ is similarly determined by appealing to the definition~\eqref{K_def} with the constant $\mu$ having a non-zero finite limit in the PN expansion, \emph{i.e.} $\mu=\Ocal(c^0)$. In particular, with eq.~\eqref{Qnr} this implies the following leading PN orders
\begin{subequations}\label{Kcalnr3}
  \begin{align}
     \Kcal &= \frac{\mu^2\Xi^2}{c^4} + \Ocal( c^{-6} )\,,
\\
     \Kcal_\Qcal &= - \frac{2\mu^2 \Xi}{c^2} + \Ocal( c^{-4} )\,,
\\
     \Qcal\Kcal_\Qcal &= - \frac{2\mu^2 \Xi}{c^2} + \Ocal( c^{-4} )\,.
\label{Kcalnr3c}
   \end{align}
\end{subequations}
The above PN orders hold for the choice we have made for the function $\Kcal$ in the action, \emph{i.e.} the $(Q-1)^2$ case. But obviously, if we include in $\Kcal$ higher powers of $Q-1$ as we do in section~\ref{Sec:cosmo-MOND} this will not change the leading PN limit~\eqref{Kcalnr3}.

\subsection{PN expansion of the field equations}
\label{PNexp}

\subsubsection{The $ij$ components of the Einstein equation}

Considering first the $ij$ components of the field equation~\eqref{EFE}, using the facts that $\Jcal$ and $\Kcal$ are small quantities $\Ocal(c^{-4})$ together with $n_i=\Ocal(c^{-1})$, we obtain\footnote{We note for reference the leading PN behaviour of the components of the Einstein tensor:
\begin{align*}
	G^{00} &= \frac{2}{c^2}\Delta\psi + \Ocal( c^{-4} )\,,
	\\
	G^{0i} &= \frac{2}{c^3}\left[ \Delta\zeta^i -  \grad^i \grad_j \zeta^j + \grad^i\dot{\psi}\right] + \Ocal( c^{-5} )\,,
	\\
	G^{ij} &= \frac{1}{c^2}\left[ \Delta\left(\phi-\psi\right)\gamma^{ij} - \grad^i \grad^j  \left(\phi-\psi\right)\right] + \Ocal( c^{-4} )\,.
\end{align*}
}
\begin{equation}
\Delta\left(\phi-\psi\right)\gamma_{ij} 
- \grad_i \grad_j \left(\phi-\psi\right) = \Ocal(c^{-2})\,.
\label{EFEij}
\end{equation}
where $\Delta \equiv \gamma^{ij} \grad_i \grad_j$ is the Laplace operator. The previous equation yields, in the case of regular isolated matter systems,
\begin{equation}
\phi = \psi + \Ocal( c^{-2})\,.
	\label{phi_psi_equality}
\end{equation}
The equality of the two potentials $\phi$ and $\psi$ in the nonrelativistic limit is very important for the viability of the theory as an alternative to dark matter, as it implies that the light deflection and the gravitational lensing is given by the same formula as in GR. That is, for any nonrelativistic baryonic distribution of matter for which the forces $\grad\phi$ are in harmony with observations as if dark matter was present, the same matter distribution will also lead to the correct gravitational lensing signal, again as if dark matter was present. This fact is actually true for the general class of modiﬁed Einstein-Aether theories (in particular the $\AEST$ theory) and was first noticed in~\cite{ZFS07}.

\subsubsection{The $00$ Einstein equation}

Next we investigate the equation satisfied by the potential $\phi$. This follows from the 00 component of the Einstein field equations~\eqref{EFE}. In this case we find that the Khronon kinetic term $\Kcal$ in the action contributes to the nonrelativistic limit through the term $\propto c^2\Qcal\Kcal_\Qcal$. Using eq.~\eqref{Kcalnr3c} and $\psi=\phi$ we obtain
\begin{equation}\label{eqphi}
	\Delta\phi + \grad \cdot \left(\Jcal_\Ycal \,\grad \Xi\right) + \mu^2 \,\Xi  = 4\pi G \rhoM + \Ocal( c^{-2})\,.
\end{equation}
\subsubsection{The $0i$ Einstein equation}

The equation for the gravitomagnetic potential $\zeta_i$ is found using the $0i$ Einstein equation~\eqref{EFE}, and to leading PN order it gives
\begin{equation}\label{eqzetai}
	\Delta\zeta^i - \grad^i \grad_j \zeta^j - \grad^i\dot{\phi} -\left[  \grad \cdot \left(\Jcal_\Ycal \,\grad \Xi\right)   + \mu^2 \Xi\right]  \grad^i\sigma
  = 4\pi G \rhoM v_m^i + \Ocal( c^{-2})\,,
\end{equation}
where once again we have used the equality of the potentials $\phi$ and $\psi$ through~\eqref{phi_psi_equality}.

\subsubsection{Consistency of the PN field equations}

In analogy with the matter equations~\eqref{PNTmunu} and following~\cite{Flanagan23}, we define the Khronon ``mass density''
\begin{subequations}\label{khrononrhov}
\begin{equation}
\label{khrononrho}	
\rho_\tau \equiv -\frac{1}{4\pi G}\left[ \grad \cdot \left( \Jcal_\Ycal \grad \Xi\right) + \mu^2 \Xi\right] \,,
\end{equation}
and the Khronon ``velocity'' field
\begin{equation}
\label{khrononv}	
v_\tau^i \equiv - \grad^i\sigma\,.
\end{equation}
\end{subequations}
Such definitions are motivated by the fact that the components of the Khronon stress-energy tensor~\eqref{Tmunutau} in leading PN order take the same form as for ordinary matter, \emph{i.e.} 
\begin{subequations}\label{TcalPN}
	\begin{align}\Tcal^{00} &= \rho_\tau c^2 + \Ocal( c^0) \,,\\\Tcal^{0i} &= \rho_\tau v_\tau^i c + \Ocal( c^{-1})\,, \\ \Tcal^{ij} &= \Ocal( c^0)\,.
	\end{align}
\end{subequations} 
The above PN scaling of the Khronon tensor justifies the PN expansion we have adopted for the metric~\eqref{PNmetric} and Khronon field~\eqref{tausigma}. 

With the previous definitions, we may rewrite the Einstein field equations~\eqref{eqphi} and~\eqref{eqzetai} in the useful form
\begin{subequations}
\label{EE}
	\begin{align}
\label{EEa} \Delta\phi &=  4\pi G \bigl(\rhoM + \rho_\tau \bigr) + \Ocal( c^{-2} )\,,
\\[0.2cm]
	\Delta\zeta^i - \grad^i \grad_j \zeta^j  - \grad^i\dot{\phi} &= 4\pi G \left(\rhoM v_m^i + \rho_\tau v_\tau^i\right) + \Ocal( c^{-2})\,,
\end{align}
\end{subequations}
which shows that the Einstein field equations, together with the continuity equation~\eqref{continuity} for matter, imply the continuity equation for the Khronon quantities~\eqref{khrononrhov}, \emph{i.e.}
\begin{equation}
\label{Khreq}
	\dot{\rho}_\tau + \grad_i\left(\rho_\tau v_\tau^i\right) = \Ocal( c^{-2})\,.
\end{equation}
The latter equation is nothing but the slow motion approximation of the Khronon field equation $\nabla_\mu\Scal^\mu=0$ since from eq.~\eqref{Scal} we have
\begin{subequations}
	\begin{align}
	\Scal^0 &= - 4 \pi G \,\rho_\tau + \Ocal( c^{-2} ) \,,\\
	\Scal^i &= - 4 \pi G \,\frac{\rho_\tau v_\tau^i}{c} + \Ocal( c^{-3} )\,. 
	\end{align}
\end{subequations}
and the Kronon equation reduces to $\dot{\Scal}^0+c\,\partial_i\Scal^i = \Ocal(c^{-2})$. This shows the consistency of the Khronon equation with the Einstein field equations in the slow motion approximation.

\subsubsection{Check in adapted coordinates}

By performing a coordinate transformation $\{x^\mu\}\longrightarrow\{\bar{x}^\mu\}$ such that  $\bar{t}\equiv \bar{x}^0/c =t+\sigma/c^2$ and $\bar{x}^i=x^i$, one obtains coordinates adapted to the foliation --- the so-called unitary gauge as used in appendix~\ref{Sec:adapted}. Indeed the Khronon field becomes
\begin{equation}
	\bar{\tau} = \bar{t} + \Ocal(  c^{-4} )\,,
\end{equation}
while the ADM form of the metric in these coordinates reads
\begin{subequations}\label{PNmetricadapted}
	\begin{align}
		\bar{N} &= 1 + \frac{\Xi}{c^2} + \Ocal(  c^{-4} )\,, \\
		\bar{N}_{i} &= \frac{1}{c} \grad_i \sigma + \frac{1}{c^3}\left[ 4\zeta_i + 2\phi\grad_i \sigma - \sigma \grad_i \dot{\sigma}\right] +\Ocal( c^{-5} )\,, \\
		\bar{q}_{ij} &= \gamma_{ij}\left(1 - \frac{2\psi}{c^2}\right) - \frac{1}{c^2} \grad_i \sigma\grad_j\sigma + \Ocal( c^{-4} ) \,,
	\end{align}
\end{subequations}
where we recall that $\Xi = \phi - \dot{\sigma} + \frac{1}{2}|\grad\sigma|^2$. With adapted coordinates (to leading PN order) we can directly insert~\eqref{PNmetricadapted} into the variational equations for the lapse, shift and spatial metric in appendix~\ref{Sec:adapted}. In particular we find that the momentum constraint equation~\eqref{constr} is identically satisfied to leading order ($\propto 1/c$) while the Hamiltonian equation~\eqref{ham} and evolution equation~\eqref{evol} are equivalent to the equality of potentials~\eqref{phi_psi_equality} together with the equation~\eqref{eqphi}. More work should also permit to confirm eq.~\eqref{eqzetai} by working out the momentum constraint equation~\eqref{constr} to next-to-leading order $\propto 1/c^3$.


To summarize, the MOND dynamics is recovered in the unitary gauge as well, as it should be, since physics does not depend on the gauge choice. This was the approach taken in Ref.~\cite{BM11}. However, as was realized in Ref.~\cite{Flanagan23} and we have verified in this section, using the unitary gauge is incompatible with the standard PN expansion in~\eqref{PNmetric}, and rather, the expansion of the shift vector acquires the term $\grad_i \sigma/c$ which dominates over the usual $\zeta_i/c^3$ assumed in eq.~\eqref{PNNi} which is the standard case. Doing that, one does recover the full PN limit, including the MOND dynamics, in a consistent way, however, the resulting expansion is more complicated to compute because of the $\grad_i \sigma/c$ term in $\bar{N}_i$. Ref.~\cite{BM11} used the unitary gauge but without the $c^{-1}$ term in the $\bar{N}_i$ expansion, concluding that the MOND dynamics is recovered. This would be equivalent to setting $\sigma=0$ throughout section~\ref{PNexp}, and this is consistent only in stationary systems. Outside stationarity, $\sigma$ must necessarily be included to avoid inconsistencies with solving the PN equations. This is seen through the Khronon equation~\eqref{Khreq}, which when $\sigma=0$ implies that $\dot{\phi}=0$ and from~\eqref{EEa} this latter condition is consistent with the PN expansion only if $\dot{\rho}_m = 0$, that is, for stationary systems.\footnote{In GR, one could start the expansion of $\bar{N}_i$ at order $c^{-1}$, but one then finds that the $c^{-1}$ term can be removed by a gauge transformation and so the standard PN expansion is justified there. This is also correct in extensions of GR, provided one includes perturbations for all additional fields. However, imposing gauge conditions on the additional fields (\textit{e.g.} the unitary gauge), removes the freedom to gauge away the $c^{-1}$ term in $\bar{N}_i$. Put it differently, as is relevant to our case, the same gauge transformation which removes the $c^{-1}$ term from $\bar{N}_i$, necessarily introduces a perturbation to the Khronon.}



\subsubsection{Stationary configurations}

We now consider the case where the system is, in addition, stationary. This means that $\dot{\phi}=\dot{\sigma}=0$ for the fields and $\dot{\rhoM} =\dot{v}_m^i=0$ for the matter, thus, the continuity equation for matter reduces to the constraint $\grad_i\left(\rhoM v_m^i\right)=0$. Furthermore, the Khronon equation~\eqref{Khreq} reduces to $\grad\cdot(\rho_\tau \grad\sigma)=0$ which we can solve by choosing $\sigma=0$. This is the unitary gauge, where the Khronon field has no perturbation and it is given by the time coordinate exactly. The choice of the unitary gauge then leads to $\Xi=\phi$ and the acceleration vector~\eqref{Amunr} becomes 
\begin{subequations}\label{Ainr}
	\begin{align}
		A_0 &= \Ocal( c^{-3} )\,,\\	
		A_i &= \grad_i\phi + \Ocal( c^{-2} )\,.
	\end{align}
\end{subequations}
Thus the acceleration of the congruence of unit vector field $n_\mu$ reduces to the physical Newtonian acceleration, and from~\eqref{eqphi} the equation for the Newtonian potential $\phi$ becomes the modified Poisson (or modified Helmholtz) equation
\begin{equation}\label{eqMOND}
	\grad\cdot\left[  \left(1+\Jcal_\Ycal\right)\grad\phi\right] + \mu^2 \phi  = 4\pi G \rhoM + \Ocal( c^{-2} )\,.
\end{equation}

\subsubsection{Choice of function $\Jcal(\Ycal)$}
\label{Jcal_choice}

The equation~\eqref{eqMOND} is still too general as we have not yet specified the function $\Jcal(\Ycal)$. Indeed, one can get a wide variety of solutions through such a choice, many of which may not necessarily lead to either Newtonian nor MONDian dynamics and may not fit observations altogether. Our goal is to specify the conditions that $\Jcal$ needs to satisfy in order that both Newtonian and MONDian regimes emerge under appropriate conditions. The fact that we shall recover the MOND regime only for stationary systems is still satisfying because most tests of MOND (rotation curves of galaxies, Tully-Fisher relation, \emph{etc.}) are performed in stationary situations.
 
From the first term in~\eqref{eqMOND} we readily identify the MOND interpolating function $\muf(y)$, where $y\equiv\vert\nabla\phi\vert/a_0$, as\footnote{We denote the MOND function as $\muf(y)$ instead of the usual $\mu(y)$ to avoid conflict with the constant scale $\mu$ in eq.~\eqref{eqMOND}. We may also denote $\chis(y)=f(y)-1=\Jcal_\Ycal(\Ycal)$ the ``gravitational susceptibility'' of the dark matter (in the sense of~\cite{B07mond}).}
\begin{equation}\label{mondfunction}
	\muf(y) = 1 + \Jcal_\Ycal(\Ycal)\,,
\end{equation}
and from~\eqref{Ainr} we can just take $\Ycal = A_\mu A^\mu/c^4$ to be $\Ycal \simeq a_0^2 y^2/c^4$ in the nonrelativistic limit. Admissible functions $\Jcal(\Ycal)$ are those for which the MOND function is linear, $\muf(y)\simeq y$, for low accelerations $y\ll 1$, while it tends to a constant in the strong field regime when $y\gg 1$. The latter constant can be set to be 1 so that $G$ represents the measured Newton's gravitational constant.

Integrating~\eqref{mondfunction} we obtain the required behaviour of the function $\Jcal$ in order for MOND to emerge in the limit $y \ll 1$:
\begin{equation}\label{lambdaMond}
	\Jcal = \Lambda - \Ycal + \frac{2c^2}{3 a_0}\,\Ycal^{3/2}+ \Ocal\left(y^4 / a_0^2\right)\,,
\end{equation}
where $\Lambda$ is the cosmological constant. Note that the low acceleration regime $y \ll 1$ is also the regime of linear cosmological perturbations, since the acceleration $A_\mu$ is a first order quantity in cosmological perturbations. Thus, $\Ycal$ is second order in cosmology and therefore, $\Lambda$ is the cosmological constant measured in cosmological observations, \emph{e.g.} the CMB.

With the above condition on the function $\Jcal$,\footnote{Note that $\Jcal$ is always of order $\Ocal(c^{-4})$, and $\Jcal_{\Ycal}$ of order $\Ocal(c^0)$, irrespective of the expansions~\eqref{lambdaMond} or~\eqref{lambdaGR}. The constant $a_0$ serves as a second expansion parameter, which determines corrections to MOND or to Newton, and is independent of the PN parameter $c^{-1}$.} the equation~\eqref{eqMOND} reduces to the usual MOND equation, \emph{i.e.}, having the Bekenstein-Milgrom form~\cite{BekM84}, apart from the presence of the ``mass'' term $\mu^2 \phi$. This term is absent from the BM theory, and is the result of the function $\Kcal(\Qcal)$, in the same way as in $\AEST$ theory~\cite{SZ21} (see also~\cite{Verwayen:2023sds}). As such, the $\mu^2\phi$ term is connected to having viable cosmological solutions and provides a link between the stationary slow motion approximation and cosmology, as we discuss further in section~\ref{Sec:cosmo}.

We expect the phenomenology of spherically symmetric solutions to the Khronon theory to be very similar to that in the case of the $\AEST$ theory, studied in~\cite{Verwayen:2023sds}. The effect of the $\mu^2 \phi$ term in eq.~\eqref{eqMOND} is then to induce an oscillatory behaviour for the potential $\phi$ on scales larger than the critical scale $\rC$ given by 
\begin{equation}\label{rcritical}
	\rC\sim \left( \frac{\rM}{\mu^2}\right)^{1/3}\quad\text{with}\quad\rM \sim \sqrt{\frac{G M}{a_0}}\,,
\end{equation}
and can be neglected when the distance $r$ with respect to the center of the source (a typical galaxy) is much smaller than $\rC$. Here $\rM$ denotes the MOND transition radius ($M$ being the mass of the source) at which the physical acceleration $|\grad\phi|$ equals $a_0$. The result~\eqref{rcritical} follows from eq.~\eqref{eqMOND} in the MOND regime where $1+\Jcal_\Ycal\simeq|\grad\phi|/a_0$, by investigating the scale at which the $\mu^2 \phi$ term will dominate over the first term (``$\mu$-domination'' regime~\cite{Verwayen:2023sds}). Since the goal is to describe both the Newtonian and MOND regimes of the galaxy we must ensure that $\rC\gg\rM$. With $a_0\simeq 1.2\,10^{-10}\,\mathrm{m}/\mathrm{s}^2$ the MOND critical acceleration scale and $M\sim 10^{11}M_\odot$ for the mass of the galaxy, we have $\rM \sim 10\, \kpc$ and we generously choose
\begin{align}
\mu^{-1} \gg 100\,\kpc\,.
\label{mu_est}
\end{align}
For instance we could take $\mu^{-1} \gtrsim 1\,\Mpc$, or even larger.

Finally, in order to recover the Newtonian law in the high acceleration regime $y\gg 1$, we impose that $\Jcal$ behaves as 
\begin{equation}
	\Jcal = \Lambda_\infty + \Ocal\left(\frac{a_0}{y}\right)\,,
	\label{lambdaGR}
\end{equation}
where the constant $\Lambda_\infty$ is \emph{a priori} different from $\Lambda$ in~\eqref{lambdaMond}, and can be seen to be the effective cosmological constant in the strong-field (GR) regime. With~\eqref{lambdaGR} imposed the model will in fact recover the full GR in this regime. In particular we conclude that the theory has the same parametrized post-Newtonian (PPN) limit as GR and is therefore viable in the Solar System and for binary pulsar tests; see eq.~\eqref{PPN} where $\beta=\lambda=0$, and with $\alpha=0$ in the high acceleration regime. Notice that for the choice we have made previously for the constant $\mu^{-1}$, say $\mu^{-1} \gtrsim 1\,\Mpc$, the ``mass'' term $\mu^2 \phi$ is negligible at the scale of galaxies, and is \textit{a fortiori} negligible at the much smaller scale of the Solar System, and has no incidence on the PPN parameters.

\section{Cosmology}
\label{Sec:cosmo}

\subsection{Friedmann-Lema\^{i}tre-Robertson-Walker (FLRW) cosmology}

\subsubsection{General FLRW background}

We consider first a FLRW background cosmology in the synchronous coordinate system. For simplicity, we choose $c=1$ from this section onwards. The metric is
\begin{align}
\dd s^2 = -\dd t^2 + a^2 \gamma^{\kappa}_{ij} \,\dd x^i \dd x^j\,,
\end{align}
where $t$ is the cosmic time and $\gamma^{\kappa}_{ij}$ is a spatially homogeneous and isotropic metric of spatial curvature $\kappa$. The symmetries of the FLRW spacetime also require that the Khronon field is a function of time only, that is, $\tau =\taub(t)$, so that the background value of $\Qcal$ is
\begin{align}
	\Qcalb = \dot{\taub}\,,
\end{align}
with the dot denoting derivatives with respect to cosmic time $t$, and the overbar always denoting the background FLRW value. Furthermore, the acceleration vanishes in the background, $\Ab_\mu = 0$.

Defining the Hubble parameter as $H \equiv \dot{a}/a$, the Einstein equations~\eqref{EFE} lead to the Friedmann equations 
\begin{subequations}\label{FLRW}
\begin{align}
 3  H^2 +   \frac{3\kappa}{a^2} - \Lambda &= 8 \pi G \sum_{I\not=\Kcal}\rhob_I  +  \Qcalb \,\Kcalb_{\Qcal}  -  \Kcalb \,,
\\
 - \left( 2\dot{H} + 3 H^2 + \frac{\kappa}{a^2} \right) + \Lambda
	&=  8 \pi G \sum_{I\not=\Kcal}\Pb_I
 +   \Kcalb \,, 
\end{align}
\end{subequations}
where $\Kcalb\equiv\Kcal(\Qcalb)$ and $\Kcalb_{\Qcal}\equiv\Kcal_{\Qcal}(\Qcalb)$, and as indicated the index $I\not=\Kcal$ runs over all types of ordinary matter species, \emph{i.e.}, excluding the contribution of the Khronon field. The cosmological constant $\Lambda$ is the one appearing in eq.~\eqref{lambdaMond}. The Khronon contribution makes it possible to defining the Khronon energy density and pressure respectively as
\begin{subequations}\label{rhobPb_Kcalb}
\begin{align}
\rhob_{\Kcal} &= \frac{1}{8 \pi G} \left( \Qcalb \,\Kcalb_{\Qcal} - \Kcalb \right) \,,\\
\Pb_{\Kcal} &= \frac{\Kcalb}{8 \pi G} \,.
\end{align}
\end{subequations}
Thus, on a FLRW background, the Khronon can be casted as a perfect fluid with time-dependent equation of state $w(t)$ given by
\begin{align}
w \equiv \frac{\Pb_{\Kcal}}{\rhob_{\Kcal}} = \frac{\Kcalb}{ \Qcalb \,\Kcalb_{\Qcal} - \Kcalb } \,.
\label{w_Kcalb}
\end{align}
As the equation of state $w$ is time-dependent, this also prompts the definition of the adiabatic speed of sound $\cad^2$ defined as
\begin{align}
\cad^2 \equiv \frac{\dd\Pb_{\Kcal}}{\dd\rhob_{\Kcal}} = \frac{\Kcalb_{\Qcal}}{\Qcalb\,\Kcalb_{\Qcal\Qcal}}\,,
\label{cad_Kcalb}
\end{align}
which obeys the well-known equation $\dot{w} = 3(w-\cad^2) (1+w)H$, see \emph{e.g.}~\cite{Kopp:2016mhm}.

The above casting of the Khronon field into a perfect fluid, is consistent with the Bianchi identities and energy conservation, $\dot{\rhob}_{\Kcal} + 3 H (1+w) \rhob_{\Kcal} = 0$, which can also be derived from the Khronon equation $\nabla_\mu \Scal^\mu = 0$, where~\eqref{Scal} implies $\Scalb^0=-c^2\Kcalb_{\Qcal}$ and $\Scalb^i=0$. Specifically, these equations  may be integrated once to give
\begin{align}
\Kcalb_{\Qcal} =  \frac{I_0}{a^3}\,,
\label{Kcal_I_0}
\end{align}
where $I_0$ is a constant set by initial conditions. Note that the non-integrated version of the Khronon equation with~\eqref{Scal}, leads in the FLRW background to 
\begin{align}
	\dot{\Qcalb} + 3 H \cad^2 \Qcalb = 0\,,
\end{align}
a relation which can be sometimes useful in reducing formulas.

\subsubsection{FLRW background in adapted coordinates}

Rather than the synchronous coordinate system one can also use the adapted coordinates of appendix~\ref{Sec:adapted} where $t=\tau$. In this case, one cannot set the lapse to $1$ as in the synchronous case, and the metric must take the form
\begin{align}
\dd s^2 = -\Nb^2\dd\tau^2 + a^2 \gamma^{\kappa}_{ij} \dd x^i \dd x^j\,,
\end{align}
where $\Nb = \Nb(\tau) = 1 / \Qcalb(\tau)$ is at this point unspecified. This leads to the Friedmann equations
\begin{subequations}\label{Fried_N}
\begin{align}
3\frac{\mathcal{H}^2}{\Nb^2} + \frac{3\kappa}{a^2} - \Lambda + \Kcalb + {\Nb}\,\Kcalb_{N}  &= 8\pi G \sum_{I\not=\Kcal}\rhob_I \,,\\
-\frac{1}{\Nb^2}\Bigl[ 2 \mathcal{H}' + 3 \mathcal{H}^2 \left( 1 -2 \cad^2\right)\Bigr]  - \frac{\kappa}{a^2} + \Lambda - \Kcalb  &= 8\pi G \sum_{I\not=\Kcal}\Pb_I \,, 
\end{align}
\end{subequations}
where we pose $\mathcal{H}=\frac{a'}{a}=\Nb H$, and the dependence of $\Kcalb$ on $\Nb$ is via the condition $\Qcalb = 1/\Nb$. The Khronon equation~\eqref{Scal} in the background is equivalent to $\Nb'=3\mathcal{H}\cad^2\Nb$, and yields
\begin{align}
\Nb^2 \Kcalb_{N} = - \frac{I_0}{a^3}\,,
\label{find_N_a}
\end{align}
which becomes an algebraic relation determining $\Nb(a)$. This in turn, is used in~\eqref{Fried_N} turning it into a modified Friedmann equation and removing also the fluid interpretation  of the Khronon (recall that $I$ runs over all types of matter species but excludes the Khronon).

\subsubsection{Approximate dust solutions}
\label{app_dust}

Specifying a function $\Kcal(\Qcal)$, completely determines the form of $\rhob_{\Kcal}$, $\Pb_{\Kcal}$, $w$ and $\cad^2$. In addition, eq.~\eqref{Kcal_I_0} may then be inverted to find a solution for $\Qcalb(a)$, and equivalently~\eqref{find_N_a} may be inverted to find $\Nb(a)$. This then, when inserted into~\eqref{rhobPb_Kcalb}, determines the exact dependence of the Khronon energy density and pressure in terms of the scale factor $a$. Thus the form of $\Kcal$ fully determines the type of cosmological evolution that ensues, and, indeed, depending on this form, one can have cosmologies which are not necessarily compatible with observations in the absence of a dark matter component. Examples are, the stiff-fluid case where $\Kcal \propto \Qcal^2$ ($w=1$) and the radiation-type case $\Kcal \propto \Qcal^4$ ($w=1/3$).

Our interest is to determine the conditions on $\Kcal(\Qcal)$ that may lead to approximate dust solutions for the Khronon, in accordance with observations. Let us first say that exact dust solutions, \textit{i.e.} $w=0$, are impossible as they would imply that $\Kcal = 0$. A simple case emerges following~\cite{Scherrer2004, Arkani2004}. In the so-called shift-symmetric $k$-essence model, when the action for a scalar field $\tau$ takes the form $(X+1)^2$ with $X = (\nabla \tau)^2$ the usual canonical kinetic term, there are approximate dust solutions in cosmology. Since $X = -\Qcal^2$ for the Khronon, it is clear that approximate dust solutions exist if $\Kcal(\Qcal)$ is expandable as a Taylor series around $\Qcal=1$. This was the motivation for the specific choice made in eq.~\eqref{K_def}.

We first focus on this specific case, that we refer to as the ``quadratic'' function, which we recall here:
\begin{align}
	\Kcal(\Qcal)=\mu^2(\Qcal-1)^2\,.
	\label{choiceKcal}
\end{align}
Then $\Kcal_{\Qcal}  =  2 \mu^2 (\Qcal - 1)$ and eq.~\eqref{Kcal_I_0} can be solved for $\Qcalb$ to get
\begin{align}
	\Qcalb = 1 + \frac{I_0}{2\mu^2a^3}\,,
	\label{Qcal_I_0}
\end{align}
where $I_0$ is a constant. Hence we find that the energy density of the Khronon is
\begin{equation}\label{epsilonFLRW}
	\rhob_{\Kcal} = \frac{\mu^2}{8\pi G} \left(\Qcalb^2 - 1\right) = \frac{I_0}{8\pi G a^3}\left( 1 + \frac{\wt_0}{a^3}\right)\,,
\end{equation}
where we posed $\wt_0  = \frac{I_0}{4\mu^2}$. The constant $I_0$ is directly related to the Khronon energy density today $\rhob_{\Kcal,0}=\frac{I_0}{8\pi G}(1+\wt_0)$ (\textit{i.e.} when $a=1$); in particular it is positive meaning also that $\Qcalb \geqslant 1$. We can rewrite
\begin{equation}\label{epsilonFLRW2}
	\rhob_{\Kcal} = \frac{\rhob_{\Kcal,0}}{a^3}\,\frac{1 + \frac{\wt_0}{a^3}}{1+\wt_0} = \frac{\rhob_{\Kcal,0}}{a^3}\left( 1 - w_0 + \frac{w_0}{a^3}\right)\,,
\end{equation}
where we have also introduced the value of the Khronon equation of state today $w_0 = \frac{\wt_0}{1 + \wt_0}$, see~\eqref{wFLRW}. If the Khronon is to give full account of the total dark matter in this model the Khronon energy density today should be
\begin{equation}
\rhob_{\Kcal,0} = \frac{3 H_0^2}{8\pi G}\,\Omega_{\Kcal,0}\,,
\end{equation}
where $H_0$ is the Hubble-Lemaître parameter and $\Omega_{\Kcal,0}\sim 0.26$ is the measured fraction of dark matter. From~\eqref{epsilonFLRW2}, the Khronon energy density is approximately that of dust, \emph{i.e.} up to some additional correction scaling as $a^{-6}$, while the equation of state evolves as
\begin{equation}\label{wFLRW}
	w = \frac{\Qcalb - 1}{\Qcalb + 1 } = \frac{\wt_0}{\wt_0 + a^3}\,.
\end{equation}
In the deep past, $a\rightarrow 0$ and $w\rightarrow 1$ so that the Khronon with the quadratic functional form of $\Kcal$ behaves as a stiff fluid.
The adiabatic speed of sound is likewise given as
\begin{equation}\label{cad2FLRW}
	\cad^2 = 1 - \frac{1}{\Qcalb} = \frac{2\wt_0}{2\wt_0 + a^3}\,.
\end{equation}
The equation of state $w$ for the quadratic potential will be shown in Fig.~\ref{Fig_w} together with the results from other potentials investigated in section~\ref{Sec:cosmo-MOND}.

\subsection{Linear perturbations on a FLRW background}
\label{Sec:GDM}

We now consider linear perturbations on a FLRW background and show that one can recast the Khronon equations into the generalized dark matter (GDM) model~\cite{Hu:1998kj, Kopp:2016mhm}. This is a model defined only on the FLRW background and linearized perturbation level (and thus without a unique non-linear completion), and determined by three parametric functions: the equation of state $w(t)$, the sound speed $\cs^2(t,k)$ and viscosity $\cvis^2(t,k)$; the last two parameters appear only at the linearized perturbed level and may depend on the wavenumber $k$.

Keeping the synchronous coordinate system for the background FLRW dynamics, and considering only scalar modes, the metric is
\begin{align}
	\dd s^2 =& -\left(1  + 2\phi\right) \dd t^2 - 2 \grad_i \zeta \, \dd t \,\dd x^i + a^2 \bigl[ \left(1 - 2\psi \right) \gamma_{ij}^\kappa + \hat{D}_{ij} \nu \bigr]\dd x^i \dd x^j \,,
	\label{g_linear_FLRW}
\end{align}
where $\hat{D}_{ij} \equiv \grad_i \grad_j - \frac{1}{3} \gamma^\kappa_{ij}\grad^2 $ is a traceless covariant derivative operator ($\grad_i\gamma^\kappa_{jk} \equiv 0$). We also perturb the Khronon field to first order as
\begin{align}
	\tau = \taub + \sigma\,,
\end{align}
leading to $\Qcalb=\dot{\taub}$ as in the previous section, and $\Qcal = \Qcalb\left(1 - \phi\right) + \dot{\sigma}$, while the normal $n_\mu$ to the spacetime foliation and acceleration $A_\mu$ components are found as 
\begin{subequations}
	\begin{align}
		n_0 &=  -1-\phi\,, & n_i &= -\frac{1}{\Qcalb} \grad_i \sigma \,,
		\\
		A_0 &= 0\,, & A_i &= \grad_i  \Upsilon \,,\label{accfirstorder}
	\end{align}
\end{subequations}
respectively, where we have posed
\begin{align}\label{upsilon}
	\Upsilon \equiv \phi -  \partial_t\!\left(\frac{\sigma}{\Qcalb}\right) = \phi -  \frac{1}{\Qcalb} \Bigl(\dot{\sigma} + 3H\cad^2  \sigma\Bigr) \,.
\end{align}
From the perturbation variables for both the metric and Khronon one can define in a usual way three independent gauge invariant variables as
\begin{subequations}\label{gaugeinv}
	\begin{align}
		\Psi &\equiv \psi + \frac{1}{6} \grad^2 \nu + H \left(\zeta + \frac{a^2}{2}\dot{\nu}\right)\,,
		\\
		\Phi &\equiv \phi - \partial_t\!\left(\zeta + \frac{a^2}{2}\dot{\nu}\right)\,,
	\end{align}
\end{subequations}
together with $\Upsilon$ for the Khronon, which is directly gauge invariant since the acceleration~\eqref{accfirstorder} vanishes in the background. 

With these, we determine the  perturbed Khronon equation~\eqref{Scal} as 
\begin{align}
\grad^2\left( \dot{\Upsilon} + H \Upsilon \right)	= 4\pi G a^2 (1+w)\rhob_\Kcal \left\{\partial_t \left[  \frac{1}{\cad^2}  \left(   \phi -  \frac{\dot{\sigma}}{ \Qcalb} \right) + 3\psi  \right] 
	-  \frac{1}{a^2} \grad^2 \left(\zeta  - \frac{\sigma}{\Qcalb} \right)\right\}
	\,.
	\label{perturbed_Khronon}
\end{align}
This equation is gauge invariant, as it can be entirely expressed in terms of the gauge invariant variables $\Psi$, $\Phi$ and $\Upsilon$. In the case of an ordinary matter species ``\textit{I}'', we find the stress-energy tensor components as
\begin{subequations}\label{T_pert}
	\begin{align}
		T^{\phantom{I}\!0}_{I\phantom{0}\!0} &= -\rhob_I \bigl(1 + \delta_I\bigr)\,,
		\\[0.2cm]
		T^{\phantom{I}\!0}_{I\phantom{0}\!i} &= -\bigl(\rhob_I + \Pb_I\bigr)  \grad_i \theta_I\,,
		\\[0.2cm]
		T^{\phantom{I}\!i}_{I\phantom{i}\!0} &= -\frac{1}{a^2}\bigl(\rhob_I + \Pb_I\bigr)  \grad^i \left(\zeta-\theta_I\right)\,,
		\\[0.2cm]
		T^{\phantom{I}\!i}_{I\phantom{0}\!j} &=  \rhob_I \bigl(w_I  + \Pi_I\bigr) \delta^i_{\phantom{i}j} +  \bigl(\rhob_I +\Pb_I\bigr)  \hat{D}^i_{\phantom{i}j} \Sigma_I\,,
	\end{align}
\end{subequations}
where the density contrast is defined as $\delta_I \equiv \delta \rho_I / \rhob_I$ and where $\theta_I$ is the velocity divergence, $\Pi_I$  the pressure contrast and $\Sigma_I$ the anisotropic stress. Applying these definitions to the stress-energy tensor~\eqref{Tmunutau} of the Khronon field,\footnote{Although we can call it $T_\Kcal^{\mu\nu}\equiv\cal{T}^{\mu\nu}$ to be consistent with~\eqref{T_pert}--\eqref{Khrononvalues}, at linear order in perturbation the Khronon stress-energy tensor also includes the contribution from the function $\Jcal$ through the term $\propto\nabla_\rho \left( \Jcal_\Ycal  A^\rho \right)$ in~\eqref{Tmunutau}, where we can use $\Jcal_\Kcal=-1$ with this approximation, see eq.~\eqref{lambdaMond}.} we find 
\begin{subequations}\label{Khrononvalues}
	\begin{align}
		\delta_{\Kcal} &=  \frac{1+w}{\cad^2} \left( \frac{\dot{\sigma}}{\Qcalb}  - \phi\right) + \frac{\grad^2\Upsilon}{4\pi G a^2\rhob_{\Kcal}}\, ,
		\label{delta_Kcal}
		\\
		\theta_{\Kcal} &= \frac{\sigma}{\Qcalb} \, ,
		\label{theta_Kcal}
		\\
		\Pi_{\Kcal} &=  \bigl(1 + w\bigr) \left( \frac{\dot{\sigma}}{\Qcalb} - \phi\right) \, ,
		\label{Pi_Kcal}
		\\
		\Sigma_{\Kcal} &= 0 \, .
	\end{align}
\end{subequations}
Combining the above results for $\delta_{\Kcal}$ and $\theta_{\Kcal}$ with eq.~\eqref{upsilon} we note that the perturbation variable $\Upsilon$ for the Khronon can be obtained by solving the equation
\begin{align}\label{equpsilon}
	\grad^2\Upsilon =  4\pi G \rhob_{\Kcal} a^2\left\{\frac{1+w}{\cad^2}\Upsilon + \Delta_{\Kcal}\right\} \,,
\end{align}
where we have introduced the (necessarily gauge-invariant) co-moving density contrast of the Khronon as
\begin{align}\label{DeltaK}
 \Delta_{\Kcal} \equiv  \delta_{\Kcal} + 3 H (1+w) \theta_{\Kcal}\,.
\end{align}

From eqs.~\eqref{Khrononvalues}, the Khronon fluid is no longer an adiabatic fluid in first order perturbation, and we have $\Pi_{\Kcal} = \cad^2 \,\delta_{\Kcal} + \Pinad$ where the non-adiabatic pressure is 
\begin{align}
 \Pinad &=  -\frac{\cad^2  }{ 4 \pi G a^2 \rhob_{\Kcal} } \grad^2  \Upsilon\,.
\label{eq_Pinad}
\end{align}
Furthermore, after further calculation, one can show that 
\begin{align}
\Pinad = \left(\cs^2 - \cad^2\right) \Delta_{\Kcal}\,,
\label{eq_Pinad_Delta}
\end{align}
where $\cs^2(t,\vec{x})$  is the speed of sound (generally a function of both time and space) and $\Delta_{\Kcal}$ is the co-moving density contrast~\eqref{DeltaK}. The relation~\eqref{eq_Pinad_Delta} is non-trivial and does not in general hold for any stress-energy tensor of the type \eqref{T_pert}; when it does (as in our case), the stress-energy tensor takes the form of the GDM model. Specifically, the speed of sound is found to be (in Fourier space)\footnote{From eq. \eqref{equpsilon} we have $\Upsilon=-\frac{\cs^2}{1+w}\Delta_{\Kcal}$ in Fourier space.}
\begin{align}
 \cs^2(t,k) = \cad^2\left[1 + \frac{\cad^2 k^2}{ 4 \pi G a^2 \rhob_{\Kcal} (1 + w)}\right]^{-1}\,,
\label{cs_Kcal}
\end{align}
and so as $k\rightarrow 0$, we have $c_s^2 \rightarrow \cad^2$, that is, the Khronon is adiabatic in the limit $k\to 0$. 

Turning to the Einstein equations~\eqref{EFE}, perturbed to linear order about the FLRW background, it is convenient to define the variable $\eta \equiv \psi + \frac{1}{6}  \grad^2  \nu $. With the previous definitions, they are given, see \textit{e.g.} eqs.~(2.13) in~\cite{Kopp:2016mhm}, by the two constraint equations 
\begin{subequations}
	\label{constraint}
	\begin{align}
		- 3  H \bigl(\dot{\psi} + H\,\phi\bigr)  
		+ \frac{1}{a^2} \left[ \left(\grad^2 + 3 \kappa \right) \eta + H \grad^2 \zeta
		\right] 
		&= 4 \pi G \sum_I \rhob_I \delta_I \, ,
		\\
		\dot{\eta}  +  H \phi + \frac{\kappa}{2} \left(  \dot{\nu} +   \frac{2}{a^2}   \zeta\right) 
		&= 4 \pi G  \sum_I   \bigl(\rhob_I + \Pb_I\bigr)   \theta_I \, ,
	\end{align}
\end{subequations}
that are found through the components $G^0_{\phantom{0}0}$ and $G^0_{\phantom{0}i}$ of the Einstein tensor, and the two propagation equations found by spliting $G^i_{\phantom{i}j}$ into a trace part 
\begin{subequations}
	\label{propagation}
	\begin{align}
		\ddot{\psi} +   3 H \dot{\psi} +  H \dot{\phi} +  \left[2\dot{H}  + 3 H^2  \right] \phi 
		+ \frac{1}{3a^2}  \grad^2 \left[  \phi - \eta  -  \dot{\zeta} -   H \zeta  \right]
		- \frac{\kappa}{a^2}\eta
		= 4 \pi G \sum_I \rhob_I \Pi_I  \,,
\end{align}
and the traceless part 
\begin{align}
		\ddot{\nu} + 3 H \dot{\nu}  +  \frac{2}{a^2} \left( \eta -\phi + \dot{\zeta} + H \zeta \right) = 16 \pi G \sum_I \bigl(\rhob_I +\Pb_I\bigr)\Sigma_I \,.
\end{align}
\end{subequations}
In eqs.~\eqref{constraint}--\eqref{propagation} the sums over index ``\textit{I}'' include the Khronon field $I=\Kcal$ through the definitions~\eqref{Khrononvalues} together with the background values in section~\ref{app_dust}. Furthermore, switching to the fluid variables for the Khronon, turns~\eqref{perturbed_Khronon} into the fluid continuity equation for the density contrast
\begin{align}
 \dot{\delta}_\Kcal 
=
3 H (w - \cs^2) \delta_\Kcal 
-9H^2 (1+w)( \cs^2 - \cad^2)  \theta_\Kcal
- \frac{1+w}{a^2}\grad^2 \left(\zeta - \theta_{\Kcal}  \right)
+ 3  (1+w)\dot{\psi} \,,
\end{align}
while a fluid Euler equation is found by combining the expression~\eqref{delta_Kcal} with the time derivative of~\eqref{theta_Kcal} as
\begin{align}
  \dot{\theta}_\Kcal  =   \cs^2  \left(3H \theta_{\Kcal} + \frac{\delta_{\Kcal}}{1+w} \right) + \phi = 
 3 H \cad^2 \theta_{\Kcal} +\frac{\Pi_{\Kcal}}{1+w} + \phi
\,.
\end{align}

The above equations for $\delta_{\Kcal}$ and $\theta_{\Kcal}$ are the standard shearless fluid equations. Thus we conclude, up to linear order around FLRW, that the Khronon behaves as a GDM fluid~\cite{Hu:1998kj,Kopp:2016mhm} with zero viscosity $\cvis^2=0$, non-zero sound speed $\cs^2$ in perturbations given by \eqref{cs_Kcal}, and time-dependent equation of state $w = w(a)$. We note, however, that once second order or higher perturbations are included, the correspondence with a perfect fluid will be lost. Moreover, the Khronon field does not contain a vector mode perturbation (being a scalar), thus, it does not lead to a purely vectorial velocity component even within the linearized fluid appoximation, in sharp contrast with standard CDM.

For any function $\Kcal(\Qcal)$ in the action for which $w\rightarrow 0$ and $\cad^2 \rightarrow 0$, in the sense that they are arbitrarily small throughout the history of the Universe for $a\gtrsim 10^{-5}$ (\emph{i.e.} for which the background solution to the Khronon energy density is approximately that of dust), the linear perturbations of the Khronon also behave as linear perturbations of dust. Thus, the approximate dust solutions discussed in~\ref{app_dust} can be extended to the linearized regime, so that the Khronon model is expected to fit large scale structure and CMB data equally well as the $\Lambda$-CDM concordance model. Specifically, we can conservatively use constraints on GDM taken from~\cite{Kopp:2018zxp, Ilic:2020onu} and apply them to the Khronon model, however, due to the $k$-dependent sound speed which doesn't fall under the models studied in~\cite{Kopp:2018zxp, Ilic:2020onu}, a proper comparison of the Khronon model with data is left for a future investigation.

\subsection{Khronon cosmology and the MOND limit}
\label{Sec:cosmo-MOND}

\subsubsection{Tension between the cosmology and MOND in the case of the quadratic function}
\label{tension}

As discussed above, with any choice for the function $\Kcal(\Qcal)$ the Khronon field behaves like GDM on a FLRW background plus linear perturbations. Furthermore, the functional choice $\Kcal=\mu^2(\Qcal-1)^2$, the ``quadratic'' function, leads to approximate dust solutions in the late universe which  can make the Khronon in agreement with cosmological observations, notably the CMB and large scale structure, provided these dust solutions can be extended as far back as the radiation-matter equality. However, the equation of state~\eqref{wFLRW} for the quadratic function is not exactly zero but tends to $w\rightarrow 1$ in the Early Universe. The turning point may be found from~\eqref{wFLRW} and is around $\wt_0 \sim  a^3$.\footnote{From~\eqref{wFLRW} the inflection point of the curve $w(a)$ occurs at $\wt_0 = 2a^3$.} The constraints on GDM found in~\cite{Kopp:2018zxp, Ilic:2020onu} require that $w\lesssim 0.0164$ around $a\sim 10^{-4.5}$ at the $99\%$ level (see the yellow shaded region in Fig.~\ref{Fig_w}) with smaller values required until $a\sim 10^{-2}$ after which the constraints become weaker. It is thus sufficient to set
\begin{equation}
	\wt_0 \approx w_0 < 5.3\times 10^{-16}  \qquad \mbox{(for cosmology)}\,,
	\label{wt_cosmo}
\end{equation}
which represents the equation of state of the Khronon today. Now, from our discussion below~\eqref{Qcal_I_0}, we have $I_0 \approx 8\pi G \,\rhob_{\Kcal,0}$ and $\wt_0  = I_0/(4\mu^2)$ so that combining the two (and restoring the relevant factors of $c$) we find
\begin{equation}
	\wt_0 =  \frac{2\pi G \,\rhob_{\Kcal,0}}{\mu^2c^2} = \frac{3 H_0^2 \,\Omega_{\Kcal,0}}{4\mu^2c^2}\,.
	\label{w_0_mu}
\end{equation}
Given that from observations $H_0 \sim 70 \,\km /s/\Mpc \sim 2.33 \times 10^{-4} \,c/\Mpc$ and $\Omega_{\Kcal,0} \sim 0.26$ we find that $\wt_0 \sim  1.06 \times 10^{-8} (\mu^{-1}/\Mpc)^2$ and thus cosmology places the bound
\begin{equation}
	\mu^{-1}  \lesssim 0.22 \,\kpc  \,.
	\label{w_0_constraint}
\end{equation}
Such a bound is incompatible with having a MOND limit in galaxies, as even for our own galaxy, we should have MOND behaviour out to tens of $\kpc$. Indeed, reversing the problem and requiring 
\begin{equation}
\label{requireMOND}
\mu^{-1} \gtrsim 100\,\kpc\,,
\end{equation}
a very optimistic bound coming from matching the MOND phenomenology,\footnote{The requirement may in fact be even larger than~\eqref{requireMOND} when matching to real data [see eq.~\eqref{mu_est}].} leads to the constraint
\begin{equation}
	\wt_0 \gtrsim 1.06 \times 10^{-10} \qquad \mbox{(for galaxies)}\,.
	\label{wt_galaxies}
\end{equation}
Clearly then, these two inequalities on $\wt_0$ given by~\eqref{wt_cosmo} and~\eqref{wt_galaxies} are in conflict. Hence, the exact functional dependence for $\Kcal$ chosen in~\eqref{choiceKcal} cannot be in simultaneous harmony with observations of galaxies and with cosmology. Below, we investigate how different choices of the function $\Kcal(\Qcal)$ may save the situation.

\subsubsection{Case of functions with higher (cubic and quartic) powers}

From the discussion in~\cite{Scherrer2004}, what we generally need in order to have dust solutions in the late Universe is that $\Kcal$ be expandable as a Taylor series around $\Qcal=1$, that is, admissible functions should be expandable as
\begin{align}
	\Kcal = 2\mu^2 \sum_{n=1}^{+\infty} \frac{\Kcal_{n+1}}{n+1} \left(\Qcal -1 \right)^{n+1} \,,
	\label{Taylor}
\end{align}
where $\Kcal_{n+1}$ are  dimensionless constants and $\Kcal_{2} = 1$ to match our convention in~\eqref{choiceKcal}.

To study the cosmological behaviour for a single power indexed by $n$, let us consider the functional form $\Kcal = \frac{2\mu^2}{n+1} \Kcal_{n+1} (\Qcal -1 )^{n+1}$ for integer $n\geqslant 2$. Then from~\eqref{Kcal_I_0} we find that
\begin{align}
	\Qcalb   =&1 + \left( \frac{I_0}{ 2\mu^2 \Kcal_{n+1} a^3}  \right)^{1/n}\,,
\end{align}
so that the energy density and pressure are found from~\eqref{rhobPb_Kcalb} to be
\begin{align}
	\rhob_{\Kcal} &= \frac{I_0}{8 \pi G \,a^3} \left[ 1 +  \frac{n}{n+1}   \left( \frac{I_0}{2 \mu^2 \Kcal_{n+1} a^3}  \right)^{1/n}  \right]  \,,
	\\
	\Pb_{\Kcal} &= \frac{I_0}{8 \pi G (n+1) a^3} \left( \frac{I_0}{ 2\mu^2 \Kcal_{n+1} a^3}  \right)^{1/n}\,,
\end{align}
and the equation of state is
\begin{align}
	w =& \left( \frac{I_0}{ 2\mu^2 \Kcal_{n+1} \,a^3}  \right)^{1/n}  
	\left[ n+1 + n   \left( \frac{I_0}{ 2\mu^2 \Kcal_{n+1} a^3}  \right)^{1/n} \right]^{-1}\,.
\end{align}

Clearly then, in the early universe, as $a\rightarrow \infty$, the Khronon once again behaves approximately as dust, however, the next-to-leading order behaviour is different. Importantly, for large $a$, the equation of state $w$ scales as $a^{-3/n}$ and, hence, for $n>1$, this powerlaw is shallower than for the quadratic case, allowing us a better chance to reconcile the phenomenology of galaxies (MOND) with cosmology. Moreover, in the PN limit the relevant quantity to check (see section~\ref{PNexp}) is
\begin{align}\label{PNok}
	c^2\Kcal_\Qcal = \frac{2\mu^2 \Kcal_{n+1} ( - \Xi)^{n}}{c^{2n-2}}\,,
\end{align}
and specifically for $n=2$ we get that $c^2\Kcal_\Qcal \sim \Xi^2 / c^2$ --- which represents a 1PN correction negligible at the level of eq.~\eqref{eqphi}. Thus for any $n\geqslant 2$ the function $\Kcal$ does not contribute to the lowest PN limit and so we have more freedom to choose the coefficients $\Kcal_{n+1}$ in order to pass the cosmological constraints~\eqref{wt_cosmo}. In the extreme case where we disallow the $n=1$ term altogether, no conflict occurs between the MOND limit in galaxies and cosmology.
\begin{figure}[!t]
\includegraphics[width=\textwidth]{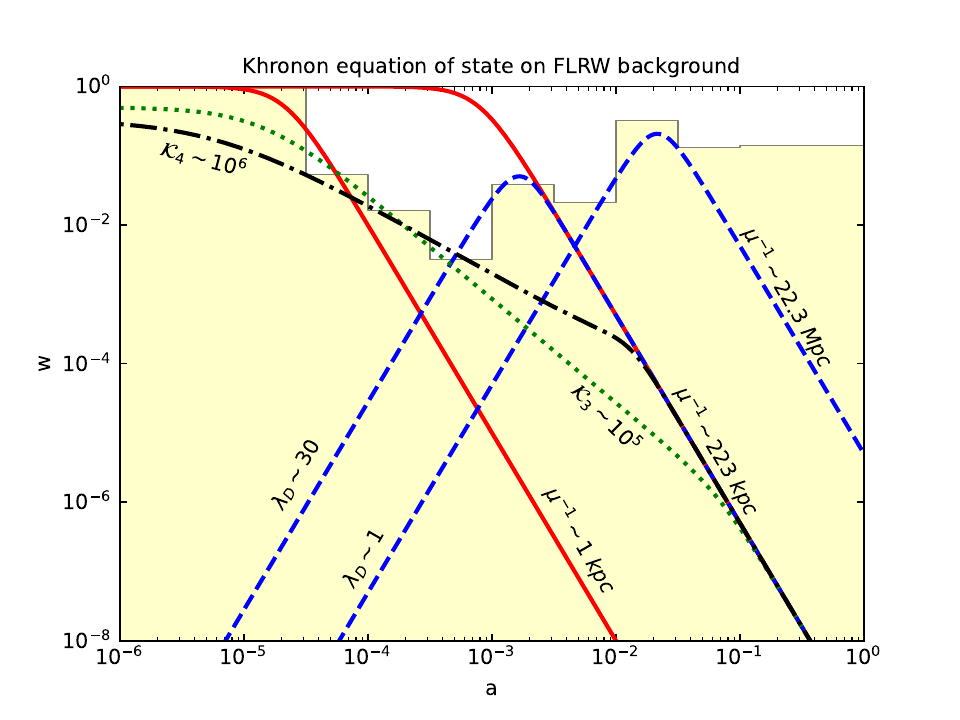}
\caption{The Khronon equation of state parameter $w(a)$ versus scale factor $a$ for several models. Displayed are two quadratic models $\Kcal=\mu^2(\Qcal-1)^2$ (solid red curves), one with $\mu^{-1} \sim 223\,\kpc$ which is consistent with a MOND limit but which cannot fit the CMB, and the other with $\mu^{-1} \sim 1\,\kpc$ which can fit the CMB but is inconsistent with a MOND limit. Introducing a cubic term with $\Kcal_3 \sim 10^5$ (green dot) or a quartic term with $\Kcal_4 \sim 10^6$ (black dash-dot) makes $\mu^{-1} \sim 223\,\kpc$ compatible with the CMB. The two DBI-inspired models~\eqref{DBIfunction} with $\mu^{-1} \sim 223\,\kpc$ and $\lambdad\sim30$, and $\mu^{-1} \sim 22.3\Mpc$ and $\lambdad\sim1$ (blue dashed) are compatible with both a MOND limit and the CMB. The yellow shaded region is allowed by the data: it corresponds to the $99\%$ credible regions of the binned ``\textit{var-w}'' model of~\cite{Kopp:2018zxp, Ilic:2020onu}, using only Planck Power Spectrum data,
and with the equation of state piecewise parametrized into eight redshift bins. 
}
\label{Fig_w}
\end{figure}

Realistically, though, it is unlikely that $n=1$ can be ignored as, unless fine-tuning is present, typical functional forms will always have the $n=1$ term (\textit{i.e.}, the quadratic function) in the expansion~\eqref{Taylor}. In that case, we expect the Universe to go through different phases where $w$ scales initially as $a^{-3}$ (the $n=1$ case) and then starts to gradually turn to different scalings until it settles to a final value in the early Universe. With this, we can expect to reconcile galaxies with cosmology, and two specific examples are shown in Fig.~\ref{Fig_w}: one case combining $n=1$ and $n=2$ with $\Kcal_3 \sim 10^5$, and one case combining $n=1$ and $n=3$ with $\Kcal_4 \sim 10^6$ (both cases assuming $\mu^{-1} \sim 223\,\kpc$). Unfortunately, we find that the required numerical values for $\Kcal_3$ or $\Kcal_4$ are unnaturally large. More generally, it seems that if the expansion~\eqref{Taylor} terminates at a finite order, large numbers for $\Kcal_{n+1}$ are inevitable, which warrants our next investigation below.

\subsubsection{Dirac-Born-Infeld (DBI) type of functions}
\label{Sec:DBI}

We would like to explore a function which has a well-defined Taylor expansion at $\Qcal=1$ in order to reproduce the dust solutions we have explored in the previous sub-sections, see eq.~\eqref{Taylor}, but which tends to a different behaviour when $\Qcal \gg 1$. A particular case is the Dirac-Born-Infeld (DBI) inspired function~\cite{Born:1933pep,Born:1934gh,Dirac:1962iy} of the form
\begin{align}\label{DBIfunction}
	\Kcal(\Qcal) = \frac{2\mu^2}{\lambdad} \left[1 - \sqrt{1 - \lambdad\left(\Qcal-1\right)^2 }  \right]\,,
\end{align}
where $\mu$ is the usual parameter as above and $\lambdad$ is a new dimensionless parameter. At $\Qcal=1$ we have $\Kcal = 0$ and more specifically, for small $\Qcal-1$ it admits the expansion $\Kcal \sim \mu^2 (\Qcal-1)^2 + \cdots$ which matches the quadratic function~\eqref{choiceKcal} and thus reproduces the required dust phenomenology in the late Universe; furthermore, as we have seen in~\eqref{PNok}, the PN limit is still safe. More generally, using~\eqref{Kcal_I_0}, we find that
\begin{align}\label{QDBI}
	\Qcalb  = 1 +  \frac{1}{\sqrt{ \lambdad + \left(\frac{2\mu^2a^3}{I_0}\right)^2}}\,.
\end{align}
Hence, as $a\rightarrow \infty$ we have $\Qcalb\rightarrow 1$ (late Universe) while as $a\rightarrow 0$ then $\Qcalb \rightarrow 1 + \lambdad^{-1/2}$ (early Universe). Thus $\Qcalb$ is bounded through $1 \leqslant \Qcalb  <  1 + \lambdad^{-1/2}$.

The energy density and pressure are found from~\eqref{rhobPb_Kcalb} to be
\begin{subequations}\label{rhoPb_DBI_a}
\begin{align}
	\rhob_{\Kcal} &=  \frac{1}{8\pi G}\left(\frac{I_0}{a^3} + \frac{2\mu^2}{\lambdad}\left[ -1 + \sqrt{1 + \lambdad \left(\frac{I_0}{2\mu^2a^3}\right)^2} \right]\right) \,,
	\label{rhob_DBI_a}\\
	\Pb_{\Kcal} &=  \frac{1}{8\pi G}\frac{2\mu^2}{\lambdad}\Biggl[ 1 - \frac{1}{\sqrt{1 + \lambdad \left(\frac{I_0}{2\mu^2a^3}\right)^2}}\Biggr] \,,
	\label{Pb_DBI_a}
\end{align}
\end{subequations}
respectively. By expanding~\eqref{rhob_DBI_a} we find
\begin{subequations}
\begin{align}
	\rhob_{\Kcal} &= \frac{I_0}{8\pi G a^3} + \Ocal\left(\frac{1}{a^{6}}\right)\qquad\qquad\qquad~\text{(when $a\to\infty$, late universe)}\,,\\
	\rhob_{\Kcal} &= \frac{I_0}{8\pi G a^3}\Bigl(1+\lambdad^{-1/2}\Bigr) + \Ocal\left(a^0\right)\qquad\text{(when $a\to 0$, early universe)}\,.
\end{align} 
\end{subequations}
The constant $I_0$ is determined by the Khronon energy density today (for given values of the constants $\mu$ and $\lambdad$), which is in turn related to the measured fraction of dark matter:
\begin{equation}
	\rhob_{\Kcal,0} = \frac{1}{8\pi G}\left(I_0 + \frac{2\mu^2}{\lambdad}\left[ -1 + \sqrt{1 + \lambdad \left(\frac{I_0}{2\mu^2}\right)^2} \right]\right) = \frac{3 H_0^2}{8\pi G}\,\Omega_{\Kcal,0}\,.
\end{equation}
Defining the function of the scale factor
\begin{equation}
	q(a) \equiv \frac{1}{\sqrt{ \lambdad + \left(\frac{2\mu^2a^3}{I_0}\right)^2}}\,,
\end{equation} 
\textit{i.e.} $q=\Qcalb-1$, such that $0 \leqslant q  <  \lambdad^{-1/2}$, we find from eqs.~\eqref{rhoPb_DBI_a} that the exact evolution of the equation of state and adiabatic sound speed as functions of the scale factor are
\begin{subequations}
\begin{align}
w &= \frac{ \lambdad  \Qsmall^2+ (1+ \Qsmall) \sqrt{1-\lambdad  \Qsmall^2} - 1 }{2+ (1+\lambdad)\Qsmall}\,,\label{w_DBI_a}\\
\cad^2 &= \frac{ \Qsmall (1 -\lambda  \Qsmall^2)}{1+ \Qsmall}\,.	
\end{align}
\end{subequations}
Alternatively it is interesting to pose $\Zcal(a)\equiv\frac{I_0}{2\mu^2 a^3}=q(1-\lambdad q^2)^{-1/2}$ and the equation of state and adiabatic sound speed become
\begin{subequations}
	\begin{align}
		w &= \frac{\Zcal}{ 1 + \lambdad \Zcal^2  + (1 + \Zcal) \sqrt{  1 + \lambdad \Zcal^2  } }\,,\\
		\cad^2 &=
		\frac{  \Zcal  } {
			\left(1 +  \lambdad \Zcal^2  \right) \left[ \sqrt{1 +  \lambdad \Zcal^2 }    +  \Zcal 
			\right]
		}\,.	
	\end{align}
\end{subequations}

The great improvement of the DBI model with respect to the previous ``$\Kcal_2+\Kcal_{n+1}$'' models, in that this model behaves approximatively as dust not only today but also in the early universe, hence it will easily pass the constraints from the CMB. In the late Universe, when $a$ is large, we may expand the above relations to get 
\begin{subequations}\label{wlateearly}
\begin{align}\label{wlate}
	w = \frac{I_0}{4\mu^2 a^3} + \Ocal\left(\frac{1}{a^{6}}\right)\,,\qquad \cad^2 = \frac{I_0}{2\mu^2 a^3} + \Ocal\left(\frac{1}{a^{6}}\right)\,,
\end{align}
while in the early Universe, when $a$ is small, they expand as
\begin{align}\label{wearly}
	w = \frac{2\mu^2 a^3}{I_0\sqrt{\lambda_D}(1+\sqrt{\lambda_D})} + \Ocal\left(a^{6}\right)\,,\qquad \cad^2 = \frac{4\mu^4 a^6}{I_0^2\lambda_D(1+\sqrt{\lambda_D})} + \Ocal\left(a^{12}\right)\,,
\end{align}
\end{subequations}
respectively. The turning point between the two regimes depends on the parameters $\mu$ and $\lambdad$. For fixed $\mu$, increasing $\lambdad$ moves the turning point to later times and \textit{vice-versa}. Increasing $\mu$, generally moves the turning point to earlier times. This behaviour may be crudely seen by equating the early and late expansions of $w(a)$ above as a first approximation; the precise relations can be found numerically, leading to monotonic dependence of $a_\text{turn}$ as a function of the parameters.

The equation of state for the DBI like function is plotted in Fig.~\ref{Fig_w} (blue dashed curves) for the two values $\mu^{-1}\sim 22.3\,\Mpc$ and $\mu^{-1}\sim 223\,\kpc$ that are compatible with the MOND phenomenology, \textit{i.e.} for which the constraint~\eqref{requireMOND} is satisfied. As we see from Fig.~\ref{Fig_w} the DBI function can be made compatible with the CMB as well, with reasonable values of the parameter $\lambdad$. For instance the DBI function with $\mu^{-1}\sim 22.3\,\Mpc$ and $\lambdad\sim 1$ fulfills well the purpose of matching both MOND and the correct cosmology and CMB anisotropies. We note, however, that the constraints from~\cite{Kopp:2018zxp, Ilic:2020onu} do not directly apply, as they correspond to purely time-dependent sound speed and viscosity, while in our case the sound speed is also $k$-dependent. In this sense, our choice of using the ``\textit{var-w}'' model is the most conservative with regards placing rough bounds on the present models. A proper comparison of the Khronon model with data is left for a future investigation.

\section{Linear stability on Minkowski space}
\label{Sec:stability}
 
\subsection{The normal modes}

We now turn to the question of the linear stability of small fluctuations on a Minkowski background. We thus expand the metric as
\begin{align}
	g^{\mu\nu} = \eta^{\mu\nu} - h^{\mu\nu}\,,
\end{align}
and the Khronon field around a background value $\bar{\tau} = t$, as
\begin{align}
	\tau = t + \sigma\,,
\end{align}
so that $\Qcal = 1 -  \phi + \dot{\sigma}$. Hence, only the quadratic part of the function $\Kcal$ in the limit $\Qcal\rightarrow 1$ will contribute to this order in the action, that is, one can consider only $\Kcal(\Qcal) = \mu^2 \left(\phi - \dot{\sigma}\right)^2$. Meanwhile $A_i = \grad_i \left( \phi -  \dot{\sigma}\right)$ so that $\Ycal = |\grad\left(\phi -  \dot{\sigma} \right)|^2$. Hence, to lowest order the MOND term does not contribute. In order to be general enough, let us set $\Jcal = -\alpha \Ycal + \cdots$ where $\alpha$ is a constant which in the MOND limit equals $1$ and in the Newtonian limit equals $0$ (in section~\ref{Sec:Horava} we discuss the more general case related to the khronometric Ho{\v{r}}ava gravity). Given this, we find that the action~\eqref{K_action} when expanded to second order in fluctuations reads
\begin{align}
S &=  \frac{1}{16\pi G}\int \dd^4x \bigg\{ - \frac{1}{2} \nablab_\mu h \nablab_\nu h^{\mu\nu}
	+ \frac{1}{4}  \nablab_\rho h \nablab^\rho h
	+ \frac{1}{2} \nablab_{\mu} h^{\mu\rho}  \nablab_\nu h^\nu_{\;\;\rho}
	- \frac{1}{4} \nablab^\rho h^{\mu\nu}  \nablab_\rho  h_{\mu\nu}
\nonumber
\\
&\qquad\qquad + 2 \alpha |\grad\left(\dot{\sigma}  - \phi\right)|^2  
	+ 2 \mu^2 (  \dot{\sigma}  - \phi )^2
	+  8\pi G \,T_{\mu\nu} h^{\mu\nu}   
	\bigg\}\,.
	\label{second_order_action}
\end{align}
Notice that since the only new terms compared to GR are the ones containing the term $\dot{\sigma}  - \phi$, that is, only a scalar mode, both tensor modes and vector modes behave exactly like in GR. The former propagate at the speed of light while the latter are non-dynamical.

We thus focus on the scalar mode action. We take the metric from~\eqref{g_linear_FLRW} and set $a=1$ and $\kappa=0$ for a Minkowski background as well as setting the matter source to zero (i.e. $T_{\mu\nu}= 0$) leading to the scalar mode action as\footnote{In terms of the gauge-invariant variables~\eqref{upsilon}--\eqref{gaugeinv} (with $a=1$) we have
\begin{align*}
	S = \frac{1}{8\pi G}\int \dd^4x \bigg\{ 
		-3 \dot{\Psi}^2 + |\grad\Psi|^2 + 2 \Phi \grad^2\Psi + \alpha |\grad\Upsilon|^2  
		+ \mu^2 \Upsilon^2
		\bigg\}\,.
\end{align*}
}
\begin{align}
S &= \frac{1}{16\pi G}\int \dd^4x \bigg\{ 
	6 \left( \frac{1}{6} \grad^2\dot{\nu}     - \dot{\psi} \right) \left( \frac{1}{6} \grad^2\dot{\nu}     + \dot{\psi} \right)
	+ 4 \left(  \frac{1}{6} \grad^2\dot{\nu}  +  \dot{\psi} \right) \grad^2\zeta
	+ 2 |\grad\psi|^2
	- \frac{2}{3}\psi \grad^4\nu
\nonumber
\\
&\qquad\quad + 4\left( \grad^2\psi  + \frac{1}{6} \grad^4\nu\right) \phi
	+ \frac{1}{18}  |\grad (\grad^2\nu)|^2
	+ 2 \alpha |\grad\left(\dot{\sigma}  - \phi\right)|^2  
	+ 2 \mu^2 (  \dot{\sigma}  - \phi )^2
	\bigg\}\,.
	\label{scalar_mode_action}
\end{align}
Choosing the Newtonian gauge for which $\zeta = \nu = 0$ (and $\psi=\Psi$, $\phi=\Phi$) and switching to Fourier space
by defining the Fourier variables via
\begin{align}
\Ocal(\vec{x})  = \sqrt{8\pi G} \int \frac{\dd^3k}{(2\pi)^3} e^{i\vec{k} \cdot \vec{x}} \Ocal_{\vec{k}}\;\;,
\end{align} 
for any operator $\Ocal$,
leads to
\begin{align}
	S &= \int \frac{\dd t \,\dd^3k}{(2\pi)^3} \left[
	-3 | \dot{\psi}_{\vec{k}}|^2 
	+  k^2 |\psi_{\vec{k}}|^2
	-  k^2 \left(  \phi_{\vec{k}} \psi_{\vec{k}}^* + \psi_{\vec{k}}  \phi_{\vec{k}}^*  \right)
	+  \left(\alpha k^2 +  \mu^2\right) \left|\dot{\sigma}  - \phi\right|^2
	\right]\,.
\end{align}
To find the normal modes we let all variables be proportional to $\de^{\di\omega t}$ and, defining the vector in field space through $\Wcal \equiv \{ \psi, \phi, \sigma \}$, the above action is written as
\begin{align}
	S =  \int \frac{\dd t\,\dd^3k}{(2\pi)^3} \; \Wcal^\dagger \Mcal \Wcal\,,
\end{align}
where $\Mcal$ is the matrix
\begin{align}
\Mcal &= 
\begin{pmatrix}
 -3  \omega^2   +  k^2  & -k^2 &  0\\
 -k^2  &   \alpha k^2 +  \mu^2   &\di  (\alpha k^2 +  \mu^2 )   \omega\\
 0 & -\di  (\alpha k^2 +  \mu^2 )   \omega   &   (\alpha k^2 +  \mu^2 )  \omega^2 
\end{pmatrix}\,.
\end{align}
The normal modes are determined by setting $\det\Mcal=0$, leading to the condition $\omega^2 = 0$. Hence, just like GR, there are no propagating wavelike modes,
 that is modes which evolve as $\de^{\di\omega t}$ with non-zero $\omega$. However, there are two non-propagating modes with $\omega=0$, one of which is dynamical, that is, of the form $A_{\vec{k}} + B_{\vec{k}} t$
(see \cite{Dubovsky:2004sg} for the existence of such modes in generic massive gravity theories).
The constant mode $A_{\vec{k}}$ will not lead to any instability and the linearly unstable mode $B_{\vec{k}} t$ only leads to a mild instability (not exponential). By considering the Hamiltonian description below, we show that this is nothing but a Jeans instability which is only exhibed at very low momenta and is thus harmless. 

\subsection{The Hamiltonian formulation}
\label{sec:Hamiltonian}

We start again from the scalar mode action~\eqref{scalar_mode_action} before fixing the gauge, and switch to Fourier space to get
\begin{align}
	S &=   \int \frac{\dd t\,\dd^3k}{(2\pi)^3} \bigg\{ 
	- 3|\dot{\psi}_{\vec{k}}|^2 
	+   \frac{1}{12} k^4 |\dot{\nu}_{\vec{k}}|^2
	+  k^2  \left[ \left(  \frac{1}{6} k^2\dot{\nu}_{\vec{k}}  -  \dot{\psi}_{\vec{k}} \right) \zeta_{\vec{k}}^* + c.c.  \right]
	\nonumber
\\
&
+  k^2 |\psi_{\vec{k}} - \frac{1}{6} k^2 \nu_{\vec{k}} |^2
	- k^2 \left[ \left( \psi_{\vec{k}}  - \frac{1}{6} k^2\nu_{\vec{k}}\right) \phi_{\vec{k}}^*  + c.c.  \right]
	+  \left(\alpha k^2 + \mu^2\right) \left|\dot{\sigma}_{\vec{k}}  - \phi_{\vec{k}}\right|^2  
	\bigg\}\,,
	\label{scalar_mode_action_Fourier}
\end{align}
where $c.c.$ means complex conjugate of the term within the square bracket. Notice that~\eqref{scalar_mode_action_Fourier} does not contain any time derivatives of the fields $\zeta_{\vec{k}}$ and $\phi_{\vec{k}}$, hence, we expect those to lead to constraints. The canonical momenta are found as 
\begin{align}
 P^{(\psi)}_{\vec{k}} &= - 2  \left( 3  \dot{\psi}_{\vec{k}} + k^2 \zeta_{\vec{k}}  \right)\,,
	\\
 P^{(\nu)}_{\vec{k}}  &=  \frac{1}{6} k^4  \left(\dot{\nu}_{\vec{k}} +  2\zeta_{\vec{k}}   \right) \,,
	\\
 P^{(\sigma)}_{\vec{k}} &=  2(\alpha k^2 +  \mu^2 )   (  \dot{\sigma}_{\vec{k}}  - \phi_{\vec{k}} ) \,,
\end{align}
and, after a Legendre transformation, the Hamiltonian is found to be
\begin{align}
 H &= 
  \int  \frac{\dd^3k}{(2\pi)^3} \bigg\{ 
 -\frac{1}{12} \bigl|P^{(\psi)}_{\vec{k}}\bigr|^2 + \frac{3}{k^4}  \bigl| P^{(\nu)}_{\vec{k}}\bigr|^2 + \frac{1}{4(\alpha k^2 +  \mu^2 )  }  \bigl| P^{(\sigma)}_{\vec{k}}\bigr|^2  	
	\nonumber
	\\
&\qquad\qquad\qquad\quad 
-  k^2\bigl|\psi_{\vec{k}} - \frac{1}{6} k^2 \nu_{\vec{k}} \bigr|^2
+ C_\zeta  \zeta_{\vec{k}}^* + C_\phi  \phi_{\vec{k}}^* + c.c.
	\bigg\}\,,
	\label{scalar_mode_Hamiltonian_preconstrained}
\end{align}
where $C_\phi $ and $ C_\zeta $ are two constraints
\begin{subequations}\label{C_phizeta}
\begin{align}
	C_\phi &\equiv  \frac{1}{2}  P^{(\sigma)}_{\vec{k}}  +  k^2  \psi_{\vec{k}}   - \frac{k^4}{6} \nu_{\vec{k}}   \approx 0\,,
\label{C_phi}
	\\
	C_\zeta &\equiv -  P^{(\nu)}_{\vec{k}} - \frac{k^2}{6} P^{(\psi)}_{\vec{k}}  \approx 0\,,
\label{C_zeta}
\end{align}
\end{subequations}
imposed by the non-dynamical fields $\phi$ and $\zeta$, respectively. As usual  we use the symbol $\approx$ to denote weakly vanishing constraints (those that vanish only on-shell).

We require that the constraints are preserved by time evolution according to the Hamiltonian $H = \int \frac{\dd^{3}k}{(2\pi)^3}\Ham$, with $\Ham$ being the Hamiltonian density in Fourier space. We define the Poisson brackets on phase space as 
\begin{align}
	\Poisson{f}{g}
	&=  (2\pi)^{3}\int \dd^{3}k \left[ \sum_A  \left(\frac{\delta f}{\delta X^A} \frac{\delta g}{\delta P^*_{X^A}} -  \frac{\delta g}{\delta X^A} \frac{\delta f}{\delta P^*_{X^A}} \right) \right]\,,
\end{align}
where  $X^A \equiv \{\psi_{\vec{k}}, \nu_{\vec{k}}, \sigma_{\vec{k}}\}$ and $P_{X^A} \equiv \{ P^{(\psi)}_{\vec{k}}, P^{(\nu)}_{\vec{k}}, P^{(\sigma)}_{\vec{k}}\}$. The Poisson brackets define the time evolution of a variable $f$ via $\dot{f} = \Poisson{f}{\Ham}$, and applied to the constraints~\eqref{C_phizeta} give
\begin{subequations}
\begin{align}
	\dot{C}_{\phi} & =  C_\zeta\,, 
	\\
	\dot{C}_{\zeta} &= 0\,.
\end{align}
\end{subequations}
In other words, the constraints are preserved by time evolution on-shell. Therefore as one might expect, the stability of the primary constraints in the absence of gauge fixing does not create new constraints. Having ensured the stability of constraints in the Hamiltonian, we can now simplify the system by employing gauge fixing.

In the Hamiltonian formulation, primary first-class constraints generate gauge transformations. The infinitesimal change of a phase space quantity $f$ under this gauge transformation generated by the constraint $C_{a}$ (with $a=\zeta, \phi$) is given by:
\begin{align}
	\Delta f = \Poisson{f}{C_{a}^*[\epsilon_{a}]}\,,
\end{align}
where we have introduced the smearing $C^{*}_{a}[\epsilon_{a}]$ of a constraint $C^{*}_{a}$ with test function $\epsilon_{a}$:
\begin{align}
	C^{*}_{a}[\epsilon_{a}] \equiv \int \frac{\dd^3k}{(2\pi)^{3}} \,\epsilon_{a,\vec{k}}\,C^{*}_{a,\vec{k}}\,.
\end{align}
Consider the following gauge transformations generated by the constraints $C_{\zeta}$ and $C_{\phi}$,
\begin{subequations}
\begin{align}
	\Delta \nu &= \Poisson{\nu}{ C_{\zeta}^*[\epsilon_{\zeta}]} = - \epsilon_{\zeta} \,,
	\\
	\Delta P_{\nu} &= \Poisson{P_{\nu}}{ C_{\phi}^*[\epsilon_{\phi}]} =  \frac{1}{6}k^{4}\epsilon_{\phi}\,.
\end{align}
\end{subequations}
Thus, we may set $\nu$ and $P_{\nu}$ to zero by a gauge transformation by choosing $\epsilon_{\zeta} = \nu$ and $\epsilon_{\phi}= -\frac{6}{k^{4}}P_{\nu}$. 
We then check what constraints are placed on the Lagrange multipliers $\zeta,\phi$ by this gauge fixing. We invoke two new gauge fixing constraints:
\begin{subequations}
\begin{align}
	G_{\nu}  &\equiv \nu \approx 0\,, \\
	G_{P_{\nu}} &\equiv  P_{\nu} \approx 0\,,
\end{align}
\end{subequations}
and find
\begin{subequations}
\begin{align}
	\Poisson{G_{\nu}}{\Ham} &= \frac{6}{k^4}   G_{P_{\nu}}  -  2\zeta_{\vec{k}}\,,
	\\
	\Poisson{G_{P_{\nu}}}{\Ham} &= \frac{1}{3} k^4  \left( \phi_{\vec{k}} - \psi_{\vec{k}}   \right) + \frac{1}{18} k^6 G_{\nu}\,.
\end{align}
\end{subequations}

Therefore the following gauge restrictions are placed on the Lagrange multipliers: $\zeta = 0$ and $\phi = \psi$. We recognize these conditions, respectively, as a restriction to the  Newtonian gauge and the content of the Einstein field equation here dictating equality between metric potentials in this gauge. We may adopt these conditions alongside the constraints $G_{\nu}$, $G_{P_{\nu}}$ in the Hamiltonian~\eqref{scalar_mode_Hamiltonian_preconstrained} and the primary constraints, yielding in addition
\begin{subequations}
\begin{align}
	P^{(\sigma)}_{\vec{k}}  &\approx  -2k^2  \psi_{\vec{k}} \,, 
	\\
	P^{(\psi)}_{\vec{k}}  &\approx 0\,,
\end{align}
\end{subequations}
so that the deconstrained Hamiltonian density is
\begin{align}
	\Ham^{({\rm dec})} = \int \frac{\dd^3k}{(2\pi)^3} \frac{(1 - \alpha) k^2 -  \mu^2}{ \alpha k^2 +  \mu^2 }  k^2 \; |\psi_{\vec{k}}|^2\,.
	\label{scalar_mode_Hamiltonian_deconstrained}
\end{align}
In the Newtonian case $\alpha=0$, hence, the Hamiltonian is positive for large momenta, that is for $k\geqslant\mu$.  Thus, the linear instability, discussed in the previous section and due to the $\mu^2 (\Qcal - 1)^2$  term in \eqref{K_action}, is a Jeans-type instability occuring only when $k<\mu$. We discuss more extensibly below.

\section{Discussions}
\label{Sec:discussions}

\subsection{The  Ho{\v{r}}ava Khronometric theory}
\label{Sec:Horava}

The model we have proposed is reminiscent of, and indeed was inspired~\cite{BM11, Sanders2011} by the Ho{\v{r}}ava-Lifshitz (HL) gravity~\cite{Horava2009, Blas2009, Blas2011}. In eq.~\eqref{lambdaMond} we found that in order to account for the MONDian low acceleration regime when $\Ycal\to 0$, the function $\Jcal(\Ycal)$ takes essentially the form $\Jcal = - \Ycal + \Ocal(\Ycal^{3/2})$, so that the gravitational part of the Lagrangian in adapted coordinates, see eq.~\eqref{adapted_action}, is
\begin{align}\label{LMOND}
	\Lcal_g =  \frac{c^4}{16\pi G} \sqrt{q} N \Bigl[ \Rcal + K_{ij}K^{ij} - K^{2}
	+ 2 \Ycal + \Ocal(\Ycal^{3/2}) 
	\Bigr] \,,
\end{align}
where we recall that $\Ycal=A_i A^i/c^4$ in adapted coordinates, and we skip other terms like $\Kcal(1/N)$ for this discussion. The theory~\eqref{LMOND} is only invariant under the subgroup of diffeomorphisms leaving invariant the preferred time foliation. In particular the term $2\Ycal$ implies a breaking of the local Lorentz invariance (LLI). 

On the other hand the HL gravity postulates a more general Lagrangian involving three arbitrary constants $\lambda$, $\beta$ and $\alpha$ labeling LLI-violating terms such that
\begin{align}\label{HLG}
	\Lcal^\text{HL}_g =  \frac{c^4}{16\pi G} \sqrt{q} N \Bigl[ \Rcal + (1-\beta)K_{ij}K^{ij} - (1+\lambda)K^{2}
	+ 2\alpha\Ycal + \cdots 
	\Bigr] \,.
\end{align}
The term $2\alpha \Ycal$ was added in~\cite{Blas2009, Blas2011} to provide stability of the scalar degree of freedom, within the so-called ``non-projectable'' version of the theory for which the lapse $N$ depends not only on time as in the original ``projectable'' version~\cite{Horava2009}, but also on space coordinates. In particular the constant $\alpha$ must satisfy~\cite{Blas2009}
\begin{equation}\label{alphaconstraint}
	0 < \alpha < 1 \,,
\end{equation}
while a non-zero $\beta$ leads to a speed of the tensor gravitational wave (GW) different from $c$.
 In addition to the explicit terms in~\eqref{HLG}, the ellipsis contain terms which are of high order (fourth and sixth) in spatial derivatives but with no time derivatives, and which ensure the power-counting renormalizability of the theory at high energy; these terms are suppressed below some high energy scale. Introducing the Khronon (which is the Stückelberg field associated with the broken diffeomorphism invariance), the covariant version of~\eqref{HLG} reads
\begin{align}
	\Lcal^\text{HL}_g =  \frac{c^4}{16\pi G} \sqrt{-g} \biggl[ R - \frac{1}{3}(\beta+3\lambda)\Theta^2 - \beta \sigma_{\mu\nu}\sigma^{\mu\nu}
	+ 2\alpha\Ycal + \cdots 
	\biggr] \,,
\label{HL_extension}
\end{align}
where $\Theta=K=\nabla_\mu n^\mu$ is the expansion and $\sigma_{\mu\nu}=K_{\mu\nu}-\frac{1}{3}\Theta q_{\mu\nu}$ is the shear of the spacetime congruence $n_\mu$, with $q_{\mu\nu} \equiv g_{\mu\nu}+n_\mu n_\nu$ and the extrinsic curvature $K_{\mu\nu} = \nabla_\mu n_\nu + c^{-2} n_\mu A_\nu$ (symmetric in $\mu\nu$ in the case of a hypersurface-orthogonal congruence). The theory~\eqref{HL_extension} is also recognized as a sub-class of Einstein-Aether theories~\cite{JM00, JM04} for which the Aether $n_\mu$ is hypersurface-orthogonal. 

Taking the model~\eqref{HLG}--\eqref{HL_extension} at face, the effective Newton's constant (as measured in a Cavendish type experiment) in the low energy limit, \textit{i.e.} setting the dots to zero, reads
\begin{equation}\label{GN}
	\GN = \frac{G}{1-\alpha}\,,
\end{equation}
while the PPN parameters of the theory are the same as for GR, with the exception of the preferred-frame parameters $\alpha_1$ and $\alpha_2$ given by~\cite{Blas2011, BlasPPN2011} (see~\cite{Bonetti2015} for further discussions)
\begin{subequations}\label{PPN}
	\begin{align}
		\alpha_1 &= - \frac{8(\alpha-\beta)}{1-\beta}\,,\\ 
		\alpha_2 &= \frac{(\alpha-\beta)\bigl[2\alpha(1+\beta+2\lambda) -\beta(3+\beta+3\lambda)-\lambda\bigr]}{(1-\alpha)(1-\beta)(\beta+\lambda)}\,.
	\end{align}
\end{subequations}
Furthermore the cosmological equations (on a FLRW background) are 
\begin{subequations}\label{HL_FLRW}
	\begin{align}
		3  \Bigl(1+\frac{\beta}{2}+\frac{3\lambda}{2}\Bigr) H^2 +   \frac{3\kappa}{a^2}  &= 8 \pi G \sum_{I}\rhob_I  \,,
		\\
		- \Bigl(1+\frac{\beta}{2}+\frac{3\lambda}{2}\Bigr) \left( 2\dot{H} + 3 H^2\right) - \frac{\kappa}{a^2} &=  8 \pi G \sum_{I}\Pb_I \,.
	\end{align}
\end{subequations}
The term $2\alpha\Ycal\propto A_iA^i$ in the action corresponds to a linear perturbation of the FLRW background and has no effect on~\eqref{HL_FLRW}.

The present model~\eqref{LMOND} looks to be a particular case of the HL gravity. However there is the crucial difference that in the HL model the violation of local Lorentz invariance is motivated by the completion of GR at high energy, while in~\eqref{LMOND} the violation of Lorentz invariance is supposed to be active only in the low energy, weak acceleration sector of the theory. In particular the value $\alpha=1$ which is required in eq.~\eqref{LMOND} seems to be incompatible with the measured Newton's constant~\eqref{GN} and the PPN parameters~\eqref{PPN}. This is because the term $2\Ycal$ is required to cancel the Newtonian gravity in the low acceleration regime and to replace it by the MOND gravity, thanks to the term $\sim \Ycal^{3/2}$ in~\eqref{LMOND}. However, there is no contradiction, as the measured value for $\GN$ happens in the high acceleration regime, where $\alpha = 0$, following the function $\Jcal(\Ycal)$ given by eq.~\eqref{lambdaGR}. Since we have also $\beta=\lambda=0$ we recover GR and in particular the same PPN limit as GR.

\subsection{Degrees of freedom and further discussion about the dispersion relation}
\label{sec:DOF}

At all instances the theory has three dynamical degrees of freedom when expanded on a Minkowski background to quadratic order. However, only two are propagating which are the usual massless tensor modes as in GR. The third degree of freedom is non-propagating (it has $\omega = 0$ and so there is no associated wave), but still dynamical as its solution is of the form $A_{\vec{k}} + B_{\vec{k}} t$. Such modes have been previously identified to exist in generic Lorentz-violating massive gravity theories~\cite{Dubovsky:2004sg}, and also exist in the case of AeST~\cite{SZ21,SZ22}, and for the ghost condensate theory~\cite{Arkani2004} in the limit when the higher-derivative interactions are ignored. In our case, there are no higher derivative interaction terms in the action, which are also quadratic in the fields (and so contribute to the usual dispersion relation corresponding to a linear wave equation). However, the higher order terms (for instance, the MOND term), will effectively lead to a non-linear propagation equation for the third degree of freedom, that is, the non-propagating $\omega=0$ mode will become propagating. We expect the full (non-perturbative) Hamiltonian analysis to reveal the same three dynamical degrees of freedom, similarly to~\cite{Bataki:2023uuy} but less complicated and can be used to study the perturbative Hamiltonian dynamics on more general backgrounds.

Let us discuss further the Hamiltonian and the linear instability found above. This is due to the ``condensate'' term $\mu^2 (\Qcal-1)^2$ which equals $\mu^2 \phi^2$ in the unitary gauge. 
 In the Newtonian case where $\alpha=0$, we have that the Hamiltonian is positive for $k>\mu$ but in the MOND case, where $\alpha=1$ it would seem that the Hamiltonian is always negative even as $k\rightarrow \infty$. However, eq.~\eqref{scalar_mode_Hamiltonian_deconstrained} is of the same form as the AeST Hamiltonian for the $\omega=0$ mode analysed in~\cite{SZ22}. There it was shown that including the MOND (non-linear) term, leads once more to a Hamiltonian bounded from below for $k\geqslant \mu$, and still unbounded from below for $k<\mu$. Thus, the same type of Jeans instability persists also at the MOND limit. Taking $\mu^{-1}\sim 22.3\,\Mpc$ as in Fig.~\ref{Fig_w} the deconstrained Hamiltonian is bounded from below for wavenumbers larger than $\sim 3 \times 10^{-31}\,\text{eV}$. 

There can be further terms which can influence the conclusions regarding the Hamiltonian. A specific case is the HL extension discussed  above, with parameters $\beta$ and $\lambda$ (we set $\beta=0$ to have GW  travelling at the speed of
light as required by observations).
Including the perturbations coming from~\eqref{HL_extension}, we find that the third mode in this case is propagating and has dispersion relation
\begin{align}
\omega^2 = \frac{\lambda \left[ (1-\alpha)   k^2  -  \mu^2\right]}{(2 + 3 \lambda) \left(\alpha  k^2+\mu^2\right)} k^2 \,,
\end{align}
which in the limit $\mu\rightarrow 0$ is consistent with~\cite{Blas2011}. Notice that the instability structure is unchanged from the $\beta=\lambda=0$ case of section~\ref{sec:Hamiltonian}, that is, we have that $\omega^2<0$ for $\sqrt{1-\alpha}\,k<\mu$.\footnote{This seems to depend on the sign of $\lambda$  and $2+3\lambda$. However, $2+3\lambda>0$ to have a well-defined Friedman equation~\eqref{HL_FLRW} and $\lambda>0$ to have $\omega^2>0$ in the $\sqrt{1-\alpha}\,k>\mu$ case.} However, in this case, the instability turns from linear to exponential. The corresponding deconstrained Hamiltonian is
\begin{align}
\Ham^{({\rm dec})} &=  \int \frac{\dd^3k}{(2\pi)^3} \bigg\{ \frac{\lambda }{4( 2  + 3 \lambda  ) }  |P^{(\psi)}_{\vec{k}} |^2
+     \frac{  (1-\alpha ) k^2  - \mu^2 }{\alpha k^2 +  \mu^2}  k^2 |\psi_{\vec{k}}|^2
\bigg\}\,,
	\label{scalar_mode_Hamiltonian_deconstrained_HL_ext}
\end{align}
which is well-defined in the limit 
 $\lambda\rightarrow 0$ and corresponds to \eqref{scalar_mode_Hamiltonian_deconstrained}. 
 Assuming that both $\alpha$ and $\lambda$ are small, the maximum instability rate $\Gamma_\text{max} = -\di \omega_\text{max}$  is $\Gamma_\text{max} = \mu \sqrt{\lambda/8}$ which we require to be smaller than $H_0$. Of particular interest is the subcase $\lambda = 2 \alpha$ for which $\omega^2 = k^2$ in the $k\gg \mu$ limit (corresponding to an equivalent case discussed in ~\cite{Blas2011}), that is, the Khronon perturbations are also luminal at high momenta. 
In this case, the PPN parameters~\eqref{PPN} are   $\alpha_1 = -8 \alpha$ while $\alpha_2= \frac{4\alpha^2}{1-\alpha} \ll \alpha_1$ which is negligible. The 
observational bounds on $\alpha_1$ impose $\alpha \lesssim 10^{-6}$, and hence, $\mu \lesssim 1000\,H_0$, corresponding to $\mu^{-1} \gtrsim 10 \,\Mpc$. 
This bound is consistent with having a MOND limit and can be weaker for smaller $\alpha_1$.

\section{Conclusion}
\label{Sec:conclusions}

We have proposed (extending previous works~\cite{BM11, Sanders2011}) a modified gravity theory to account for the phenomenology of dark matter at the scale of galaxies as summarized by the MOND formula~\cite{Milg1, Milg2, Milg3}. Although this theory cannot be considered as fundamental (there is an arbitrary function in the action which is not explained by fundamental physics), it owns a number of attractive features:
\begin{enumerate}
\item It is based on only two dynamical fields: the metric and the scalar Khronon field (plus ordinary matter fields);
\item It recovers MOND at the scale of galaxies (in the weak acceleration regime), with however the restriction to systems being stationary;
\item In the strong acceleration regime the theory agrees with GR and in particular has the same PPN limit as GR for tests in the Solar System and in binary pulsars;
\item The theory has no propagating GW with helicity $0$ or helicity $1$, so GW are the same as in GR;
\item Last but not least, it can be arbitrarily close to the $\Lambda$-CDM  cosmological model  at the level of linear cosmological perturbations, 
  where it retrieves the full observed spectrum of CMB anisotropies (for a wide range of parameters $\lambdad$ and $\mu$).
\end{enumerate}
We note the importance of checking the validity of the MOND limit in galaxies when~\eqref{lambdaMond} is taken as a low energy effective field theory expansion which comes with a new scale. A naive estimate is that the MOND term in~\eqref{lambdaMond}, that is, $M_p^2 |\grad \phi|^3/a_0 \rightarrow |\grad \phi_{c}|^3/\Lambda_0^2$ where $\phi_{c} \sim \phi / M_p$ is the canonically normalized version of $\phi$, leads to a strong-coupling scale of $\Lambda_0 \sim \sqrt{a_0 M_p}$. Given that $a_0 \sim H_0/6$ then $\Lambda_0 \sim  \text{meV}$ corresponding to $\Lambda_0^{-1} \sim 10^{-4}\,\text{m}$, which is clearly much smaller than the scale of galaxies. However, given that there are more dynamical degrees of freedom in the theory, this naive estimate is bound to change. We leave a more thorough investigation of this estimate to a future study.

We also hightlight that to next order in cosmological perturbations, the MOND terms will become relevant and the correspondence with $\Lambda$-CDM will break down. In that case, it is necessary to use $N$-body simulations to determine the non-linear cosmological large scale structure, and it would be interesting to check where (and how) quasistatic MONDian sources might emerge in such a setup. This is left for a future investigation.

\section*{Acknowledgements}

We thank Gilles Esposito-Far\`ese, Eanna Flanagan, Elias Kiritsis, Ignacy Sawicki and Leonardo Trombetta for interesting discussions. 
We have received support from the Barrande mobility programme (project number 8J21FR028). C.S. acknowledges support 
by the European Structural and Investment Funds and the Czech Ministry of Education, Youth and Sports (MSMT) (Project CoGraDS-CZ.02.1.01/0.0/0.0/15003/0000437)
and by the Royal Society Wolfson Visiting Fellowship ``Testing the properties of dark matter with new statistical tools and cosmological data''.
 
\appendix

\section{Formulation in adapted coordinates}
\label{Sec:adapted}

It is always possible to adopt a coordinate system for which the coordinate time $t\equiv x^0/c$ is equal to the Khronon field: $t = \tau(x)$. Hence in this coordinate system $\Qcal=1/N$ where $N=(-g^{00})^{-1/2}$ is the lapse. The unit vector is $n_{\mu}=(-N,\bm{0})$. Introducing the shift $N_{i}=g_{0i}$ and the spatial metric $q_{ij}=g_{ij}$, we have the usual 3+1 form for the metric,
\begin{equation}\label{ADM}
	\dd s^2 = - c^2 N^2 \dd t^2 + q_{ij}\bigl(\dd x^{i}+c N^{i}\dd t\bigr)\bigl(\dd x^{j}+c N^{j}\dd t\bigr) \,.
\end{equation}
The acceleration takes the 3-dimensional expression (with $A^0=0$)
\begin{subequations}\label{Ai} 
	\begin{align}
		A^{i} &= c^2 D^{i}\ln N\,,\\ \Ycal &= \frac{A_i A^i}{c^4} = D_{i}\ln N D^{i}\ln N\,,
	\end{align}
\end{subequations}
where $D^i=q^{ij}D_j$ (acting here on a scalar) denotes the covariant derivative compatible with the spatial metric, \emph{i.e.} $D_i q_{jk} = 0$. The covariant action~\eqref{K_action} becomes, after discarding a total divergence (with $\sqrt{-g} = N\sqrt{q}$)
\begin{align}\label{adapted_action}
	\!\!\!S =  \frac{c^4}{16\pi G} \!\int \!\dd t\,\dd^3\mathbf{x} \,\sqrt{q} N \Bigl[ \Rcal + K_{ij}K^{ij} - K^{2}
	- 2 \Jcal(\Ycal) + 2 \Kcal(\Qcal) 
	\Bigr] + S_m\bigl[\PsiM,N,N_{i},q_{ij}\bigr]\,,
\end{align}
where $\Rcal$ is the 3-dimensional scalar curvature of the 3-metric $q_{ij}$ and $K_{ij}$ is the extrinsic curvature
\begin{equation}\label{Kij} 
	K_{ij} = \frac{1}{2N}\left( \frac{1}{c} \partial_{t} q_{ij} - D_{i}N_{j} - D_{j}N_{i} \right) \,.
\end{equation}
In such adapted coordinates the Khronon field has disappeared and the independent dynamical degrees of freedom are geometrical: $N$, $N_i$ and $q_{ij}$ (and the matter fields $\PsiM$). See section~\ref{Sec:Horava} for a generalization of the action inspired by Ho{\v{r}}ava gravity.

For the variation of the minimally and universally coupled matter action we adopt a notation similar to that in~\cite{BM11}
\begin{align}\label{matter3d}
	\rho \equiv -\frac{1}{c^2\sqrt{q}}\frac{\delta S_{m}}{\delta N} \,,\qquad J^{i} \equiv \frac{1}{c\sqrt{q}}\frac{\delta S_{m}}{\delta N_{i}}\,, \qquad T^{ij} \equiv \frac{2}{N\!\sqrt{q}}\frac{\delta S_{m}}{\delta q_{ij}}\,,
\end{align}
where $\rho$, $J^{i}$ and $T^{ij}$ reduce to the usual notions of mass density, current density and spatial stresses in the case of a perfect fluid. The variation with respect to the lapse $N$ gives
\begin{equation}
	\label{ham}
	\Rcal + K^2 - K_{ij}K^{ij} - 2\Jcal + 4\Ycal\,\Jcal_\Ycal + 2\Kcal -2 \Qcal\,\Kcal_\Qcal + \frac{4}{c^2}   D_{i}\bigl(\Jcal_\Ycal A^{i}\bigr) = \frac{16\pi G}{c^2} \,\rho\,.
\end{equation}
Next, the variation with respect to $N_i$ yields the momentum constraint equation as in GR,
\begin{equation}\label{constr}
	D_{j}\left( K^{ij}-q^{ij}K \right) = - \frac{8\pi G}{c^3} J^{i}\,.
\end{equation}
Varying with respect to the spatial metric $q_{ij}$ gives 
\begin{align}
	&
	\frac{1}{c N}D_t\left(K^{ij}-q^{ij}K\right)
	+ \frac{2}{N}D_{k}\left[N^{(i}(K^{j)k}-q^{j)k}K)\right]
	+ 2 K^{ik}K^{j}_{\phantom{j}k}-K K^{ij}
	- \frac{1}{2}q^{ij}\left(K^{kl}K_{kl}+K^2\right)
	\nonumber 
	\\
	&\qquad 
	+ \Gcal^{ij}
	- \frac{1}{N} \left(D^iD^jN - q^{ij} D_kD^k N\right) 
	-  \frac{2}{c^4} \Jcal_\Ycal \,A^iA^j + \left(\Jcal-\Kcal\right) \,q^{ij} 
	= \frac{8\pi G}{c^4} T^{ij}\,,
	\label{evol}
\end{align}
where $\Gcal^{ij}=\Rcal^{ij}-\frac{1}{2}q^{ij}\Rcal$ is the 3-dimensional Einstein tensor, and we use the convenient notation $D_t\equiv\partial_t-c N^kD_k$. The trace of eq.~\eqref{evol}, combined with the constraint equation~\eqref{constr}, gives
\begin{equation}\label{evoltrace}
	- \frac{4}{c N}D_t K
	- 3 K_{ij}K^{ij} - K^2
	- \Rcal
	+ \frac{4}{N} D_iD^i N - 4\Ycal\,\Jcal_\Ycal + 6 \left(\Jcal-\Kcal\right) = \frac{16\pi G}{c^4}\Bigl( T^i_i + \frac{2 c}{N}N_i J^i \Bigr) \,.
\end{equation}
Finally, by eliminating $\Rcal$ between~\eqref{ham} and~\eqref{evoltrace} we obtain the following Poisson-like equation,
\begin{equation}\label{modPoisson}
	\frac{1}{c^2} D_{i}\left[\left(1+\Jcal_\Ycal\right)A^i\right] + \Jcal + \Ycal -\Kcal - \frac{1}{2}\Qcal\,\Kcal_\Qcal - \frac{D_t K}{c N} - K^{ij}K_{ij} = \frac{4\pi G}{c^2}\biggl(\rho + \frac{2}{c N}N_iJ^i + \frac{1}{c^2}T^i_i\biggr)\,,
\end{equation}
which is the generalization of eq. (3.13) in the BM theory.

\bibliographystyle{JHEP.bst} 
\bibliography{ListeRef_Khronon.bib}

\end{document}